\documentclass[aps,pre,preprint,amsmath,amssymb,showpacs,twocolumn]{revtex4-1}

\usepackage{dcolumn}
\usepackage{bm}
\usepackage{color}

\def\d{\delta}

\def\be{\begin{equation}}
\def\ee{\end{equation}}
\def\bea{\begin{eqnarray}}
\def\eea{\end{eqnarray}}
\newcommand{\e}{{\rm e}}
\newcommand{\half}{\mbox{$\frac{1}{2}$}}

\newcommand{\lb}{\label}
\renewcommand{\d}{{\rm d}}
\renewcommand{\i}{{\rm i}}

\usepackage{epsfig}

\usepackage{amsthm}
\usepackage{graphicx}

\usepackage{color}
\usepackage[utf8]{inputenc}
\usepackage[english]{babel}
\usepackage[titletoc,toc,title]{appendix}

\usepackage[top=2 cm,bottom=2 cm,left=2. cm, right=2. cm]{geometry}

\usepackage{dsfont}
{\nopagebreak\refstepcounter{theorem} \par{\it\ Assumption}\ \thetheorem{}. \ignorespaces\nopagebreak}%

\begin{document}


\title{Stochastic activation in a genetic switch model}




\author{John Hertz}
 \email{john.hertz@nordita.org}
\affiliation{Nordita, Stockholm, Sweden; Niels Bohr Institute and Institute for Neuroscience, University of Copenhagen, Denmark}%


\author{Joanna Tyrcha}
 \email{joanna@math.su.se}
\affiliation{ Matematiska institutionen, Stockholm University, Sweden}%

\author{Alvaro Correales}
\email{alvaro.correales@estudiante.uam.es}
\affiliation{Departamento de Matematicas, Ciudad Universitaria de Cantoblanco, s/n, 28049 Madrid, Spain}

\date{\today}

\begin{abstract}
We study a biological autoregulation process, involving a protein that enhances its own transcription, in a parameter region where bistability would be present in the absence of fluctuations.  We calculate the rate of fluctuation-induced rare transitions between locally-stable states using a path integral formulation and Master and Chapman-Kolmogorov equations.  As in simpler models for rare transitions, the rate has the form of the exponential of a quantity $S_0$ (a ``barrier'') multiplied by a prefactor $\eta$.  We calculate $S_0$ and $\eta$ first in the bursting limit (where the ratio $\gamma$ of the protein and mRNA lifetimes is very large). In this limit, the calculation can be done almost entirely analytically, and the results are in good agreement with simulations.  For finite $\gamma$ numerical calculations are generally required.  However, $S_0$ can be calculated analytically to first order in $1/\gamma$, and the result agrees well with the full numerical calculation for all  $\gamma > 1$.  Employing a method used previously on other problems, we find we can account qualitatively for the way the prefactor $\eta$ varies with $\gamma$, but its value is 15-20\% higher than that inferred from simulations.
\end{abstract}


\maketitle

\section{Introduction}

Fluctuations are intrinsic to biology because many biochemical processes involve small numbers of molecules \cite{Kepler,bress,bressjphysa}.  Advances in experimental techniques have made it possible to observe and measure these fluctuations directly \cite{Elowitzetal}. Furthermore they are not always just ``noise''; they can also have a function in enhancing the survival of an organism or species \cite{Eldar,Normanetal}.   Therefore, gaining a quantitative understanding of biological processes requires making reliable calculations of fluctuation rates, for example, of transitions between phenotypes.  These transitions are controlled by networks of genes and proteins.  Furthermore, even rare transitions can be important, because they can lead to big changes in the phenotype.  To help establish solid computational techniques for analyzing these networks and calculating the rates of such rare transitions we analyze here a minimal model of an autoregulatory network in which a protein, when bound to the regulatory DNA region of its own gene, enhances its transcription rate.  This positive feedback enables the protein to maintains a high-concentration state.  While the model is a drastic simplification of the true biochemical dynamics, we believe that the methods we use can be carried over to more complex networks relevant to a variety of biologically interesting phenomena.

In our model we assume that the transcription rate is a sigmoidal function of the protein concentration.  This can lead to bistability  -- both low- and high-concentration states can be locally stable. The switch from the low- to the high-concentration state is generally flipped by some other molecule that also promotes transcription of the mRNA and thereby production of the protein.  However, even when the concentration of the other molecule is too small to flip the switch, fluctuations may do so.  In this paper we study such transitions in our simple model, with the aim of understanding how different features of the biochemical circuitry affect the rate at which these (rare) events occur.  

Previous work has explored these features of genetic regulatory circuits, and much of it has become textbook material \cite{bress}.   In particular, simple autoregulatory circuit mechanisms are well-understood, as is one-dimensional Kramers escape from metastable states, both for cases where the dynamics are effectively diffusive and a Fokker-Planck description is adequate \cite{gardiner}, and for those in which molecule numbers must be treated as integers, where one must solve a master equation \cite{walczak,bressjphysa}.   

Here we adopt a path-integral approach to the problem, taking into account both mRNA and protein concentrations.   We are interested in the parameter regime in which the switching rate is very small and we can treat the problem by a saddle-point approximation.   The problem of finding the optimal path then reduces to solving the equations of motion for an appropriate Hamiltonian system that we will derive.  In general, these equations have to be solved numerically.   We will do this and compare the results with those of numerical simulations.

The limit in which both the protein lifetime is much longer than that of the mRNA and many proteins are translated from each mRNA copy is interesting and particularly relevant in bacteria.  Viewed on the timescale of protein degradation, the molecules produced from a single mRNA look like a simultaneous burst, hence the name ``translational bursting''.  Since the mRNAs typically have exponentially-distributed lifetimes, the number of protein molecules in a burst is also exponentially distributed \cite{ShahrezaeiSwain2008}.    We study this ``bursting limit'' and the approach to it as the ratio of mRNA to protein lifetimes goes to zero.  In the limit, the 2-dimensional mRNA-protein problem reduces to a 1-dimensional one.  Several groups have solved the time-independent Chapman-Kolmogorov equations for this and related systems and found the stationary protein number distributions.   Using both an extension of their methods and the 1-d limit of our path integral, we are able to calculate the switching rate analytically, and simulations confirm the theoretical predictions.

The paper is organized as follows:  First, we introduce the model and derive the path integral.  We then derive the Hamilton equations for the optimal switching path and describe how to solve them numerically and find the ``activation barrier'', i.e., the dominant exponential factor in the switching rate.
 Then we take a closer look at the bursting limit, showing how the mRNA concentration can be eliminated from the Hamilton equations and how the activation barrier can be evaluated analytically.  This result makes contact with the calculations mentioned above of the stationary distribution, and by extending the methods used by those authors we are able to evaluate the prefactor in the switching rate analytically.  Again, these theoretical predictions are found to agree with numerical simulations of this simplified model.  

We turn then to the general problem away from the bursting limit, where the two molecular species (protein and mRNA) both have to be kept explicitly in the calculation, which has to be done numerically. We calculate first the barrier, for a range of values of the ratio $\gamma$ of mRNA and protein degradation rates. It is possible to decompose the result in a natural way into protein and mRNA contributions, and we find that the former of these is quite insensitive to $\gamma$, while the latter falls off toward zero with increasing $\gamma$.  We find we can calculate the mRNA contribution analytically to first order in $1/\gamma$, and the results agrees quite well with the full numerical barrier calculation, even for $\gamma$ of order $1$.  Turning to the prefactor, we calculate it following a procedure due to Maier and Stein \cite{MaierStein92} and compare the results with numerical simulations.  Finally, we discuss briefly some open questions in this and related systems which our work may help to answer. 
 
To keep the logical flow of our presentation as simple as possible, some of our calculations that involve complicated algebra and/or are fairly straightforward extensions of treatments elsewhere in the literature are relegated to appendices.

\section{Model}\lb{modelsection}

We employ a minimal stochastic model for a gene whose transcription receives positive feedback from its own protein product.  Discretizing time, with $t_i = i\Delta t$, we describe the mRNA dynamics simply by a stochastic equation
\be
x_{i+1} - x_i = w_i,							\label{eq:firstxeq}
\ee
with the probability density of the jump $w_i$ given by
\bea\lb{eq:Pofw}
\rho(w_i) &=& \\ &=&[1 - (g(y_i)+\gamma x_i)\Delta t ]\delta(w_i) \nonumber \\&&+ g(y_i)\Delta t \delta(w_i-1)  \nonumber +\gamma x_i \Delta t \delta(w_i+1)		\nonumber
\eea
where $\gamma$ is the mRNA degradation rate, $y(t)$ is the protein concentration, and we take the production rate $g(y)$ to be given by a Hill function
\be
g(y) = a + g_0\frac{y^h}{y^h + K^h}.				\lb{eq:Hillfn}
\ee
Such a form can be derived under the assumption that the protein binding and unbinding from the DNA is fast in comparison with the timescales of the present problem, see, e.g., \cite{walczak}.  In the deterministic limit, Eqns (\ref{eq:firstxeq}) and (\ref{eq:Pofw}) lead to a rate equation
\be
\dot{x} = g(y) - \gamma x.						\lb{eq:xrateeqn}
\ee

For the protein dynamics, one can write an analogous stochastic kinetic equation, with the production rate proportional to the mRNA concentration: 
\be
y_{i+1} - y_i = v_i,							\label{eq:firstyeq}
\ee
where the jump probability density is
\bea\lb{eq:Pofv}
\sigma(v_i) = cx_i\Delta t \delta(v_i-1) +  y_i \Delta t \delta(v_i+1) \\
+[1 - (cx_i+y_i)\Delta t ]\delta(v_i). \nonumber
\eea		
We measure time in units of the protein lifetime, so the degradation rate is equal to $1$. 
The rate equation is
\be
\dot{y} = cx - y.								\lb{eq:yrateeqn}
\ee

\begin{figure}[ht] 
	\begin{center}
		\includegraphics[keepaspectratio=true,scale=0.45]{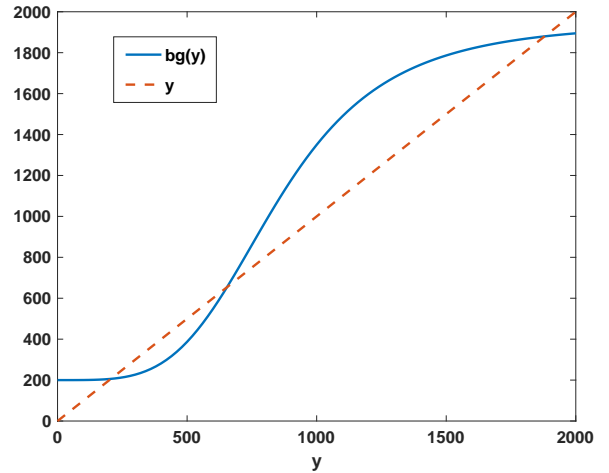}
		\caption{The blue (solid) line shows the dependence of the steady-state protein production rate and the red (dashed) line that of its degradation rate, both as functions of the protein concentration $y$. The crossing points are the metastable concentrations $y_0,y_1,y_2$}
		\label{figure1}
\end{center}
\end{figure}

In steady state, eliminating $x$ from the rate equations (\ref{eq:xrateeqn}) and (\ref{eq:yrateeqn}) gives
\be
y = \frac{c}{\gamma} g(y)  \equiv bg(y)						\lb{eq:selfconsistent}
\ee
If the Hill exponent $h > 1$, and the parameter $g_0$ in (\ref{eq:Hillfn}) is big enough, it is possible to find bistability (see Fig. \ref{figure1}): two different protein concentrations give stable solutions.  However, the fluctuations in the mRNA dynamics can cause transitions between these states.  Our aim in this paper is to study these fluctuations and, in particular, to calculate the rate of these switchings in the limit where they are rare.  This problem is thus similar to classical (two-dimensional) Kramers escape \cite{gardiner,eyring1935activated}, but the discrete molecule numbers and the higher dimensionality of the problem require new methods.

The parameter $b = c/\gamma$ appearing in (\ref{eq:selfconsistent}) is the mean number of protein molecules produced per mRNA lifetime; quite commonly $b \gg 1$.  When the mRNA lifetime $\gamma^{-1}$ is very short compared to that of the protein, the protein molecules translated from a single mRNA are effectively produced simultaneously when viewed on the protein timescale.  This kind of protein production is called ``bursting", and we will study this limit in some detail, both because of its biological relevance and to make contact with previous studies.  We will use the term ``bursting limit'' to mean $\gamma \gg 1$, even when the burst size $b$ is not large, because this is the condition that is necessary to reduce the problem to an effective one-species one.  However, the large-$b$ case is the one of biological interest.

Note also, however, that in our calculations of the rare-event rate of transitions between the phenotypes with protein concentrations near $y_0$ and $y_2$, $b$ is also the small parameter of the problem, in the sense that the activation barrier is proportional to $1/b$.  More precisely, $b$ has to be small compared to the differences $|y_1-y_{0,2}|$ between the protein concentrations in the almost-stable states and that at the unstable transition state separating them.  Thus, we require $1 \ll b \ll |y_1 - y_{0,2}|$. 

Calculating the $b$-dependence of the transition rate is our main goal.  To separate this dependence from that on other parameters of the model, we will always vary $b$ in such a way that the steady states
(\ref{eq:selfconsistent}) do not change.  This means that when mutiply $b$ by some factor, the transcription rate parameters $a$ and $g_0$ in (\ref{eq:Hillfn}) are divided by the same factor, i.e., we keep $bg(y)$ invariant.

\subsection{Path integral formulation}
Following \cite{chow}, we can go quite easily from the time-discretized stochastic differential equation (SDE) (\ref{eq:firstxeq}) to a path integral representation of the probability of a given history starting at $t=0$ and ending at $t=T$.  We start by inserting Dirac $\delta$-functions to impose (\ref{eq:firstxeq}) and (\ref{eq:firstyeq}) in the integration over every $w_i$ and using the Fourier representation of the $\delta$-functions:
\begin{widetext}
\be
P[x,y] = \int \prod_i dw_i  dv_i \frac{dp_i}{2\pi}\frac{dq_i}{2\pi}\rho(w_i)\sigma(v_i)\exp \left\{- \i
\sum_i [ p_i (x_{i+1} - x_i  - w_i)  + q_i (y_{i+1}-y_i -v_i)]\right\} .	\lb{eq:Frep}
\ee
\end{widetext}
At each time step we get factors equal to the characteristic functions for $w_i$,
\bea
& {\displaystyle \int } & dw_i\rho(w_i) \exp (\i w_i p) = \\ &=&  g(y)\Delta t \e^{\i p}+ \gamma x \Delta t \e^{-\i p} + 1 - (g(y)+\gamma x)\Delta t \nonumber
\eea
and $v_i$,
\bea
& {\displaystyle \int } &  dv_i\sigma(v_i) \exp (\i v_i q) =  \\ & = &cx\Delta t \e^{\i q}+ y \Delta t \e^{-\i q} + 1 - (cx+y)\Delta t .\nonumber
\eea
For small $\Delta t$ we can write these as
\be
\exp \left \{ - [g(y) (1-\e^{ \i p})+ \gamma x  (1-\e^{-\i p}) ]\Delta t \right \}
\ee
and 
\be
\exp \left \{ - [cx (1-\e^{ \i q})+ y  (1-\e^{-\i q}) ]\Delta t \right \},
\ee
respectively.  Then, taking the continuum limit $\Delta t \to 0$, we arrive at
\be
P[x,y]  = \int Dp Dq \exp \left(- S[x,p,y,q ]\right),					\lb{eq:Pofxandy}
\ee
where $Dp$ and $Dq$ are shorthand for the limit as $\Delta t \to 0$ of the multidimensional integrals over the $p_i$ and $q_i$ in (\ref{eq:Frep}) (including all the factors of $(2\pi)^{-1}$) and $S$, called the action, is
\begin{widetext}
\be
S[x,p,y,q ] = \int_0^T \left[ \i p \dot{x} +  g(y) (1-\e^{ \i p})+ \gamma x (1-\e^{-\i p}) +\i q\dot{y} 
+cx(1-\e^{\i q}) + y(1- \e^{-\i q}) \right]dt.
\ee
Finally, we make the shift $ip \to p$, $iq \to q$, giving
\be
S[x,p,y,q ] = \int_0^T \left[ p \dot{x} +  g(y) (1-\e^{p})+ \gamma x (1-\e^{-p}) +q\dot{y} 
+cx(1-\e^{q}) + y(1- \e^{-q})\right]dt;						\lb{eq:Sxpyq}				\ee
\end{widetext}
now the integrals in $Dp$ and $Dq$ in (\ref{eq:Pofxandy}) run along the imaginary axis.  We remark that if we expand the exponentials in (\ref{eq:Sxpyq}) to first order, we recover delta-functions leading to the noise-free rate equations (\ref{eq:xrateeqn}) and (\ref{eq:yrateeqn}).  

If we now integrate over all histories satisfying boundary conditions $x(0) = x_0$, $y(0) = y_0$, $x(T) = x_T$ and $y(T) = y_T$, we get the probability, given the initial condition, of reaching $x_T$ and $y_T$ at time $T$ by any path:
\bea\lb{eq:pathintegral}
&P&(x_T,y_T|x_0,y_0)=  \\ & =&{\displaystyle \int } Dx Dp Dy Dq \exp \left(- S[x,p,y,q ]\right)		\nonumber	
\eea
(where $x(t)$ and $y(t)$ are subject to the boundary conditions).	
The functional integration symbols $DxDp$ are defined by
\be
Dx Dp = \lim_{\Delta t \to 0} \prod_i\frac{dx_i dp_i}{2\pi} ,
\ee
where $x_i = x(t_i)$, $p_i = p(t_i)$ and $t_{i+1}-t_i = \Delta t$, and correspondingly for $Dy$ and $Dq$.

The quantity (\ref{eq:Sxpyq}) is the action for a 2-dimensional classical mechanical problem with a Hamiltonian 
\bea\lb{eq:hamilt}
H(x,p,y,q) &=& \\ & =&g(y)(\e^p - 1) - \gamma x (1 - \e^{-p}) \nonumber \\ &&+cx(\e^q-1) -y(1-\e^{-q})	\nonumber
\eea
If the noise in the problem is weak enough (we will say more specifically what this means in the present problem later), the path integral will be dominated by paths near the classical paths, i.e., the solutions of the 
Hamiltonian equations of motion
\bea
\dot{x} &=& \frac{\partial H}{\partial p} = g(y)\e^p -\gamma x \e^{-p}	\lb{eq:xdoteqn}  \\
\dot{p} &=& -\frac{\partial H}{\partial x} = \gamma (1- \e^{-p}) - c(\e^q-1),  	\lb{eq:pdoteqn} \\
\dot{y} &=& \frac{\partial H}{\partial q} = cx\e^q	 - y\e^{-q}				\lb{eq:ydoteqn} \\
\dot{q} &=& -\frac{\partial H}{\partial y} = -g'(y)(\e^p - 1) +1-\e^{-q}				\lb{eq:qdoteqn}
\eea
A problem like ours in a similar model system was studied by Assaf {\em et al} \cite{assafPRL2011}. 

In this paper, following Friedman {\em et al} \cite{fried}, we will frequently consider the case where the number of protein molecules $y$ is sufficiently large that we can treat it as continuous, with deterministic dynamics given by the rate equation (\ref{eq:yrateeqn}).  Enforcing this by means of a delta-function in computing $P[x,y]$, we have
\begin{widetext}
\be
P[x,y] = \int \prod_i dw_i  \frac{dp_i}{2\pi}\frac{dq_i}{2\pi}\rho(w_i)\exp \left\{- \i
\sum_i [ p_i (x_{i+1} - x_i  - w_i)  + q_i (y_{i+1}-y_i -(cx_i-y_i)\Delta t)]\right\} .	\lb{eq:Frepcontin}
\ee
Performing the integrations over the $w_i$ then leads to an action (after the $\i p \to p$ shift)
\be
S[x,p,y,q] =   \int_0^T \left[ p \dot{x} +  g(y) (1-\e^{ p})+ \gamma x (1-\e^{- p}) +q(\dot{y} -cx +y) 
\right]dt.									\lb{eq:Sxpyqcontin}
\ee
\end{widetext}
The Hamiltonian is
\bea \lb{eq:hamiltcontin}
H(x,p,y,q) &= & g(y)(\e^p - 1) - \gamma x (1 - \e^{-p}) \nonumber \\&&+q(cx -y),	  
\eea
with equations of motion
\bea
\dot{x} &=& \frac{\partial H}{\partial p} = g(y)\e^p -\gamma x \e^{-p}	\lb{eq:xdoteqncontin}  \\
\dot{p} &=& -\frac{\partial H}{\partial x} = \gamma (1- \e^{-p}) - cq,  	\lb{eq:pdoteqncontin} \\
\dot{y} &=& \frac{\partial H}{\partial q} = cx	 - y						\lb{eq:ydoteqncontin} \\
\dot{q} &=& -\frac{\partial H}{\partial y} = -g'(y)(\e^p - 1) +q		.		\lb{eq:qdoteqncontin}
\eea
It is simple to verify that both this Hamiltonian and its equations of motion can be obtained simply from the corresponding equations (\ref{eq:hamilt}) and (\ref{eq:xdoteqn}-\ref{eq:qdoteqn}) for the discrete-protein-number problem by expanding to first order in $q$, as one would expect for a continuum approximation.
 
Our goal will be to calculate the rate of transition from the metastable low-protein-concentration state at $y_0$ through the unstable transition state at $y_1$ to the region around $y_2$, in the parameter range where such events are rare.  This rate has the form \cite{gardiner}
\be
\Gamma = \eta \exp (-S_0).				\lb{eq:Gamma}
\ee 
The quantity $S_0$ in the exponent is the optimal or extremal value of the action $S$, obtained by
setting $\delta S/\delta x = \delta S/\delta p = \delta S/\delta y = \delta S/\delta q = 0$.  The prefactor $\eta$ comes from fluctuations around the optimal path.  In what follows, we will do this first in the limit $\gamma \gg 1$ of fast mRNA degradation, returning afterwards to the more general problem.
  
\section{Bursting limit} \lb{section:bursting}

As noted above, in the limit $\gamma \gg 1$, all the proteins translated from a single mRNA are effectively produced in a simultaneous burst when looked at on the protein degradation timescale.  Furthermore, since the mRNA lifetime is exponentially distributed, the burst size is also exponentially distributed\cite{ShahrezaeiSwain2008,bress}.  (Actually, since the protein numbers are integers, one would conventionally call the distribution geometric.)  Here we are most interested in the case where the mean burst size $b \gg 1$, so the burst number can be taken as continuous and the density to be a continuous (one-sided) exponential.  Such a distribution has been observed experimentally \cite{caifri,Yu}.

In the following two subsections, we calculate the action $S_0$ of the optimal path and the prefactor $\eta$ of (\ref{eq:Gamma}) in this limit, for both the continuous-protein-number approximate model (\ref{eq:hamiltcontin}) and the exact one (\ref{eq:hamilt}).

\subsection{Action of the optimal path}
\subsubsection{continuous protein number}
  
We start, for simplicity, in the continuous-protein-number approximation.  
If we divide both sides of the Hamilton equation  (\ref{eq:pdoteqncontin}) for $p$ by $\gamma$, the left-hand side, which is proportional to $1/\gamma$, must go to zero for $\gamma \to \infty$.  However, the individual terms on the right-hand side do not go to zero in this limit, so we obtain a condition on their sum:
\be
1- \e^{-p}=\dfrac{cq}{\gamma}=bq  					\lb{eq:gam2}.
\ee
An analogous argument can be made for Eqn. (\ref{eq:xdoteqncontin}) for $x$ (without the need to divide by $\gamma$) if we regard it as an equation for $\gamma x$.  This gives the condition
\be
g(y)\e^{2p}=\gamma x. 						\lb{eq:gam1} 
\ee

Solving for $\e^p$ and $x$ as functions of the protein variables $y$ and $q$, we find
\bea
\e^{p}=\dfrac{1}{1-bq} 			\lb{eq:epofq}
\\
x=\dfrac{g(y)}{\gamma(1-bq)^2},	\lb{eq:xofyandq}
\eea
and substituting into (\ref{eq:hamiltcontin}) yields the effective protein-only Hamiltonian
\be
H_B = q\left( \frac{bg(y)}{1-bq} - y\right) .				\lb{eq:HsubB}
\ee
The Hamilton equations of the reduced problem,
\bea
\dot{y} &=& \frac{\partial H_B}{\partial q} = \dfrac{bg(y)}{(1-bq)^2}- y  \lb{eq:ydoteqn3}
\\
\dot{q} &=& -\frac{\partial H_B}{\partial y}  = q\left( 1-\dfrac{bg'(y)}{1-bq}\right)  ,      \lb{eq:qdoteqn3}
\eea
can also be derived by using (\ref{eq:epofq}) and (\ref{eq:xofyandq}) in (\ref{eq:ydoteqncontin}) and (\ref{eq:qdoteqncontin}).

For small $bq$, we can expand $1/(1-bq)$ in (\ref{eq:HsubB}) to first order, yieding a more familiar kind of problem, with a Hamiltonian
\be
H = q[bg(y)-y] +b^2g(y)q^2.
\ee
This is the Hamiltonian associated with the Fokker-Planck equation
\be
\partial_t P(y,t) = -\partial_y [(bg(y) - y)P] + b^2 \partial_y^2 [g(y)P]
\ee
that we could derive from the Ito SDE
\be
dy = [bg(y)-y] + b\sqrt{2g(y)}dW,
\ee
where $W$ is a Wiener process.  It describes positive drift under Gaussian multiplicative noise, with a noise power $2b^2g(y)$.  In this limit, the exponential distribution of the bursts plays no role.    

It is an easy exercise to see that the Hamiltonian $H_B$ (\ref{eq:HsubB}) can be derived directly, analogously to what we did in the previous subsection, starting from the discretized SDE 
\bea
y_{i+1}-y_i=-y_i\Delta t +w_i,   \lb{eq:sde}
\eea  
with the probability of the jump $w_i$ given by
\bea \lb{eq:prob}
\rho(w_i) &=&\left[ 1-g(y_i)\Delta t \right] \delta (w_i) \\ &&+\dfrac{g(y_i)\Delta t}{b} \exp(-w_i/b),  \nonumber
\eea
expressing the fact that there is a rate of bursting equal to $g(y)$, and the burst size is exponentially distributed with mean $b$.  In this derivation, one can see that the  $1/(1-bq)$ factor in $H_B$ (\ref{eq:HsubB}) comes from the moment generating function of the exponential burst distribution.

\begin{figure}[ht] 
	\begin{center}
		\includegraphics[keepaspectratio=true,scale=0.45]{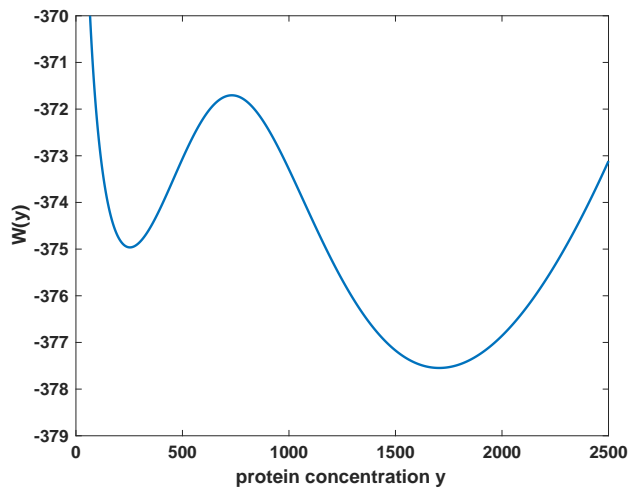}
		\caption{The effective potential $W(y)$ for the continuous-protein-number model.}
		\label{figure2}
\end{center}
\end{figure}

In terms of $H_B$ the action (\ref{eq:Sxpyq}) is
\be
S_0[y,q] =   \int_0^T \left[ q \dot{y} -H_B(y,q)\right]dt.		\lb{eq:Syq}
\ee
 The Hamiltonian is a constant of the motion, i.e., its value along any path $y(t),q(t)$ that solves the equations of motion is fixed. For the path of interest to us here, that fixed value is zero, since the path starts and ends at $q=0$. So, setting $H_B=0,$ we get either
\be
q=0  \lb{eq:q0}
\ee
or
\be
\dfrac{bg(y)}{1-bq}-y=0.        \lb{eq:qdot0}
\ee
The first possibility describes the "downhill" path from the unstable fixed point at the middle root $y_1$ of (\ref{eq:selfconsistent}) to the smaller one $y_0$. It has action $S=0$. The second possibility describes the nontrivial "uphill" path for which $q$ goes from $y_0$ to $y_1$. We can solve (\ref{eq:qdot0}) to get an explicit expression for the path:
\be
q=\dfrac{y-bg(y)}{by}.
\ee
From (\ref{eq:Syq}) and the fact that $H_B=0$, we can evaluate the action for that path as follows
\bea \lb{eq:Sol}
S_0 &=&   \int \left[ q \dot{y} -H_B(y,q)\right]dt \nonumber \\ &= &\int q\frac{dy}{dt}dt=\int_{y_{0}}^{y_{1}}q(y)dy  \\ &=&
\int_{y_{0}}^{y_{1}}\dfrac{y-bg(y)}{by}dy=W(y_1)-W(y_0)	,	\nonumber
\eea
where 
\be
W(y)=\dfrac{y}{b}-a\ln y-\dfrac{g_0}{h}\ln (y^h+K^h).  \lb{eq:WSol}
\ee 
acts as an effective potential for this model.  It is plotted in Fig. (\ref{figure2}).   The minima at $y_0$ and $y_2$ and the maximum at $y_1$ are evident.

The exponential of $-W$ is, up to normalization and a factor $1/y$, the stationary probability distribution of the process (\ref{eq:sde}). With changes in notation, this agrees with the result of Friedman {\em et al} \cite{fried}, who derived it from the Chapman-Kolmogorov equation. We also note that when we vary $b$, keeping the fixed points invariant as described above, $W(y)$ and, hence, the action are proportional to $1/b$.  Thus, the bursting size plays a role like temperature in the Arrenhius rate $\propto \e^{-S_0}$.
The condition that a description of the process in terms of rare events is valid is $S_0 \gg 1$, 
consistent with our earlier condition $y_1-y_0 \gg b$ since (\ref{eq:WSol}) is dominated by the first term for $y_{0,1} \gg 1$.
 
\subsubsection{discrete protein number}

Now let us consider the corresponding problem in the model (\ref{eq:hamilt}) with discrete protein number. Proceeding as before, we use the fact that mRNA dynamics are fast to set $\dot{x}=0$ and $\dot{p}=0$ in (\ref{eq:xdoteqn}) and  (\ref{eq:pdoteqn}):
\bea
g(y)\e^{2p}=\gamma x 						\lb{eq:gam3} \\
1- \e^{-p}=\dfrac{c}{\gamma}(e^q-1)=b(e^q-1)  \lb{eq:gam4}.
\eea
 Solving the second of these equations for $e^p$ gives
 \bea
\e^{p}=\dfrac{1}{1-b(\e^{q}-1)},			\lb{eq:epofq2}
\eea
from which we get, using (\ref{eq:gam3}),
\bea
c x=\dfrac{bg(y)}{(1-b(e^q-1))^2},	\lb{eq:xofyandq2}
\eea
and substituting into (\ref{eq:hamilt}) gives the Hamiltonian
\be
H_{Bd}= (e^q-1)\left( \frac{bg(y)}{1-b(e^q-1)} - ye^{-q}\right).				\lb{eq:HsubBd}
\ee
In this case the equations of motion are 
\bea
\dot{y} &=& \frac{\partial H_{Bd}}{\partial q} = \dfrac{bg(y)e^q}{(1-b(e^q-1))^2}- ye^{-q}  \lb{eq:ydoteqn4}
\\
\dot{q} &=& -\frac{\partial H_{Bd}}{\partial y}  = 1-e^{-q}-\dfrac{bg'(y)(e^q-1)}{1-b(e^q-1)}    .  \lb{eq:qdoteqn4}
\eea
As we expect, in the small-$q$ limit, this $H$ and these equations of motion reduce to the corresponding equations (\ref{eq:HsubB})-(\ref{eq:qdoteqn3}) for the continuous-protein-number model.  Also, in the limit of small $b$, (\ref{eq:HsubBd}) just describes a simple birth-death process with protein generation rate $bg(y)$ -- there is no bursting.  We will not consider this case further.  
 
The Hamiltonian (\ref{eq:HsubBd}) can also be derived directly starting, as before, from a discretized SDE 
\be \lb{eq:discrediscre}
y_{i+1}-y_i=w_i
\ee
with the probability of the jump given by
\bea \lb{eq:Pofw_d}
\rho(w_i) &=&   g(y_i)\Delta t(1-r)\sum_{n=0}^{\infty}r^n\delta(w_i-n) \nonumber \\ && +y_i \Delta t \delta(w_i+1) \\ &&+[1 - (y_i+g(y_i))\Delta t ]\delta(w_i) \nonumber 				
\eea
where $r=b/(1+b).$
We remark that the same quantity $b/(1+b)$ occurs in the geometric burst distribution in the paper by Shahrezaei and Swain \cite{ShahrezaeiSwain2008}.

We can now find, analogously as we did for the continuous-protein-number, the optimal path $q(y)$ from the condition $H_{Bd}=0$. From (\ref{eq:HsubBd}), in the case $q\neq 0$, we have
\be
\frac{bg(y)}{1-b(e^q-1)} = ye^{-q},				\lb{HBd0}
\ee
which can be solved for $e^{-q}:$  
 \bea
\e^{-q}=\dfrac{b(g(y)+y)}{y(1+b)},			\lb{eq:eqofHBd}
\eea
and then
 \bea
q(y)=\ln{\left(\dfrac{y(1+b)}{b(g(y)+y)}\right)}.		\lb{eq:logq}
\eea
The action for the path from $y_0$ to $y_1$ is
\be
S_0 =   \int _{y_0}^{y_1}q(y)dy				\lb{eq:SyqHBd}
\ee
Unlike the corresponding expression (\ref{eq:Sol}) for the continuous-protein-number case, this integral does not seem to be calculable analytically. Nevertheless, it is possible to expand it in $1/b$.  Writing
\be
q(y) = \ln \left( 1 + \frac{1}{b} \right) - \ln  \left( 1 + \frac{bg(y)}{by}\right),
\ee
we expand (remembering that we are taking $g(y) \propto 1/b$, so $bg(y)$ is independent of $b$) and integrate the two terms separately .  The result is
\bea \lb{eq:tayloract}
& S_0(b) =\\ =&{\displaystyle \sum_{n=1}^{\infty}}\dfrac{(-1)^{n-1}}{n}  \left( {\displaystyle \int_{y_0}^{y_1}} \dfrac{y^n-(bg(y))^n}{y^n} dy \right) b^{-n} \nonumber
\eea
The first term in the series is the expression for $S$ found above in (\ref{eq:Sol}).  For moderately large $b$, this series converges quite rapidly; for $b=15$ two terms are sufficient for $1\%$ accuracy.  Accuracy can be important here, since the mean escape time is exponential in $S$.

\subsection{Prefactors}

Let us now consider the prefactor contributions to the escape rates for our two models. While these can be calculated by expanding the expression for the action to second order in the deviations from the optimal path, we find it much simpler to go back to the fundamental stochastic descriptions (the Chapman-Kolmogorov equation for continuous protein number and the master equation for discrete protein number).

We begin with continuous bursting limit case. We start with Chapman-Kolmogorov equation for exponential bursting \cite{fried}:
\bea \lb{eq:KS}
& \partial_tP(y,t)= \\ & =\partial_y(yP(y))+\int_0^{y}w(y-y')g(y')P(y')dy'=0,	\nonumber
\eea
where, as before, we have set the protein degradation rate equal to 1,
$g(y)$ is the Hill function (\ref{eq:Hillfn}), and $w(y-y')=b^{-1}\exp[-(y-y')/b]-\delta(y-y')$, with $b$ the mean burst size.
Thus, in steady state, (\ref{eq:KS}) can be written as 
\begin{widetext}
\bea
\partial_y(yP(y))+\frac{1}{b}e^{-y/b}\int_0^{y}e^{y'/b}g(y')P(y')dy' -g(y)P(y)=0.													\lb{eq:KS2}
\eea
Now we calculate the derivative of (\ref{eq:KS2}):
\bea
\partial_y^2(yP(y))-\frac{1}{b^2}e^{-y/b}\int_0^{y}e^{y'/b}g(y')P(y')dy' +\frac{1}{b}g(y)P(y)
-\partial_y(g(y)P(y)) =0.										\lb{eg:KS3}
\eea
\end{widetext}
Adding (\ref{eq:KS2}) and $b$ times (\ref{eg:KS3}) gives
\bea
\partial_y[( y -bg(y))P(y)]+\partial_y^2{(b yP(y))}=0,
\eea
which is a steady-state Fokker-Planck equation for drift $bg(y)-y$ and multiplicative noise, i.e., a diffusion ``constant'' $D(y)$ equal to $2by$.
It is almost a standard Kramers problem to calculate the escape rate for this problem, but since easily-accessible treatments generally do not treat multiplicative noise, we present the calculation in Appendix
\ref{append:cont}. We find the following prefactor:
\bea
\eta =\frac{1}{2\pi}\sqrt{\frac{y_0}{y_1}}\sqrt{(1-bg^{'}(y_0))|1-bg^{'}(y_1)|}.\lb{eq:rate}
\eea
This differs from the additive-noise case only in the factor of $\sqrt{D(y_0)/D(y_1)} = \sqrt{y_0/y_1}$.  Note that since, when we change $b$, we hold $bg(y)$ constant, $\eta$ does not depend on $b$.

For discrete protein number, we use an analogous trick. We start with the Hamiltonian given by equation (\ref{eq:HsubBd}).  Multiplicating this equation by $1-b(e^q-1)$ gives
\begin{widetext}
\bea
\hat{H} = [1 - b(\e^q-1)]H &=& (\e^q-1)[ bg(y) - (1 - b(\e^q-1))y\e^{-q} ]			\nonumber \\ & =& (\e^q-1)[ b(g(y) +y)  - (b+1) y\e^{-q} ].	
\eea
\end{widetext}
This is a Hamiltonian for the simple birth-death process treated by Bressloff \cite{bress2} with birth and death rates
\bea
\Omega_+(y) &=& b(g(y)+y),					\lb{eq:Omegaplus}	\\
\Omega_-(y) &=& (b+1)y	,					\lb{eq:Omegaminus}
\eea
respectively, so we can simply carry over his result.  As in the continuous case, for steady state $H=0$, so we also have $\hat{H}=0$. This implies equations (\ref{eq:eqofHBd}) and (\ref{eq:logq}).  Bressloff's result (eqn (3.26) of \cite{bress2}) contains the prefactor 
\be
\eta = \frac{\Omega_+(y_0)}{2\pi}\sqrt{|S''(y_1)| S''(y_0)}. 
\ee
Now using $q(y) = S'(y)$ and differentiating (\ref{eq:logq}), gives
\be
S''(y) = \frac{1}{y} - \frac{1+g'(y)}{y+g(y)},
\ee
and because $bg(y_{0,1})=y_{0,1}$ we have
\be
S''(y_0) = \frac{1-bg'(y_0)}{(1+b)y_0}
\ee
and analogously for $S''(y_1)$. Using these results and (\ref{eq:Omegaplus}) and (\ref{eq:Omegaminus}), we get the prefactor
\bea \lb{eq:prefac11}
\eta &=& \frac{(1+b)y_0}{2\pi} \sqrt{\frac{(1-bg'(y_0))|1-bg'(y_1)|}{y_0y_1(1+b)^2}}	\nonumber
\\ &=& \frac{1}{2\pi} \sqrt{\frac{y_0}{y_1}}\sqrt{(1-bg'(y_0))|1-bg'(y_1)|}, 
\eea
identical to the continuous-protein-number result (\ref{eq:rate}). It is also possible to obtain this result using the Wentzel–Kramers–Brillouin (WKB) method.  The proof, a generalization of that of Bressloff \cite{bress2} for the simple birth-death process, is given in Appendix \ref{append:ac}.

\subsection{Simulations}\lb{sect:burstsim}

Using the Gillespie algorithm \cite{Gillespie}, we have simulated the bursting-limit models with both continuous (\ref{eq:prob}) and discrete (\ref{eq:Pofw_d}) burst-size distributions, measuring the mean time $\overline{\tau}$ to reach the unstable point $y_1$.  We take $1/(2\overline{\tau})$ as the empirical escape rate $\Gamma_{\rm sim}$, where the factor of $1/2$ comes from the fact that a system at the unstable point has a probability of $1/2$ to leave it in either direction.     We have done this for $13$ values of $b$, from $b=15$ to $b=75$ in steps of $\Delta b=5$, for both continuous and discrete protein number. For each value of $b$ we have simulated $10,000$ escape events. We find that the escape times are exponentially distributed for this range of $b$, giving an empirical error for the escape rate of $2\%$ with a confidence of $95\%$.

Throughout all our simulations, we fix the values of the parameters on the Hill function as: $a= 13.33$, $g_0 = 116.6667$ and $h=4$ and $K=850$, for $b=15$, thus preserving the fixed points $y_0 = 206.0185$ and $y_1 = 653.8648$. Whenever we vary the parameter $b$, we do it in such a way as to preserve $bg(y)$, that is, we define $a=13.33\cdot 15/b$ and $g_0 = 116.6667 \cdot 15/b$ for an arbitrary $b$. 
  
\begin{figure}[ht]
 	\begin{center}
 		\includegraphics[keepaspectratio=true,scale=0.45]{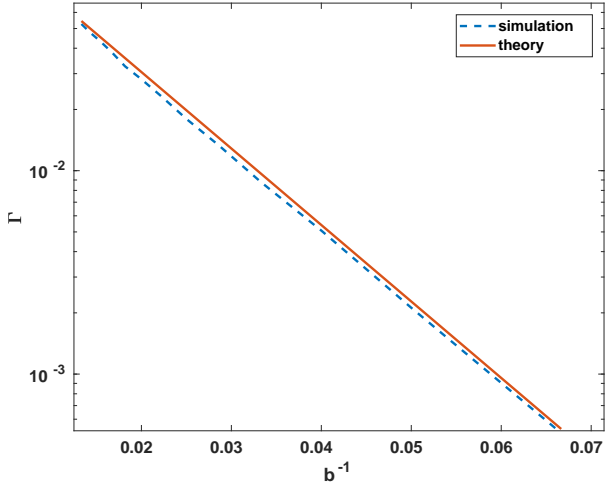}
 		\caption{Escape rate $\Gamma$ plotted against inverse burst size parameter $b^{-1}$ for the continuous-protein-number model:.  The red (solid) line shows the theoretical prediction (\ref{eq:Gamma}) and the blue (dashed) line the simulation results.}
 		 \label{figure3}
 	\end{center}
 \end{figure}

\begin{figure}[ht] 
 	\begin{center}
 		\includegraphics[keepaspectratio=true,scale=0.45]{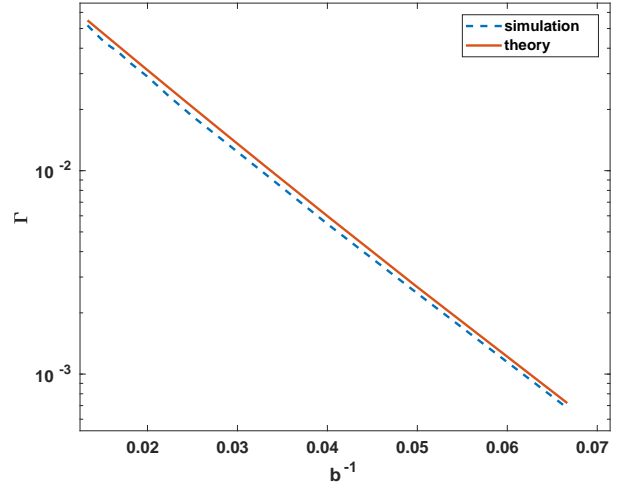}
 		\caption{Escape rate $\Gamma$ plotted against inverse burst size parameter $b^{-1}$ for the discrete-protein-number model:.  The red (solid) line shows the theoretical prediction (\ref{eq:Gamma}) and the blue (dashed) line the simulation results.}
 		\label{figure4}
 	\end{center}
 \end{figure}

The results (Figs. \ref{figure3} and \ref{figure4}) agree well with the expected form (\ref{eq:Gamma}) with the exponent $S_0$ predicted by equations (\ref{eq:Sol})-(\ref{eq:WSol}) and (\ref{eq:logq})-(\ref{eq:tayloract}) for the continuous and discrete burst-size cases, respectively, and the common prefactor given by (\ref{eq:rate}) and (\ref{eq:prefac11}).  The empirical escape rates are slightly smaller (by a few percent) than the theoretical values, but the difference shrinks as $b \to 0$.  We attribute the discrepancy to the fact that our theory is only exact in the limit of infinitesimal escape rates.  However, the simulation times necessary to confirm this quantitatively are prohibitively long.
 
It is worth mentioning that, applying the Gillespie algorithm in the continuous case, one has to take into account that degradation in protein concentration induce a time dependent rate for calculating the  probability distribution
\bea 
Prob(w(t)=0,\forall T_1<t<T_2) = \nonumber \\ =\exp \left[ -\int_{T_1}^{T_2} g(y(T_1)e^{T_1-s})ds \right],
\eea
where $w$ is the noise term in equation \eqref{eq:sde}.
 
\section{Full problem: mRNA and  protein dynamics}

\subsection{Computing the extremal action}

In the full problem described in section \ref{modelsection}, we have the extremal action in the form
\be
S_0 = \int_0^T [p_s(t) \dot{x}_s(t) + q_s(t)\dot{y}_s(t)]dt,					\lb{eq:S0full}
\ee
where $x_s$, $p_s$, $y_s$, and $q_s$ solve the Hamilton equations (\ref{eq:xdoteqn})-(\ref{eq:qdoteqn}) or (\ref{eq:xdoteqncontin})-(\ref{eq:qdoteqncontin}), since $H=0$ along the extremal path.  However, the condition $H=0$ is insufficient to determine the $x$- and $y$-dependence of $p$ and $q$ because it is only one condition on four variables.  The only recourse is to integrate the Hamilton equations to find $x_s(t)$, $p_s(t)$, $y_s(t)$, and $q_s(t)$ and then use these in (\ref{eq:S0full}). Both steps have to be done numerically.   We do this using the ``relaxation'' method described in Numerical Recipes \cite{press2007numerical}.  Details of the present calculations are summarized in Appendix \ref{append:relax}.

Figs. \ref{figure5} and \ref{figure6} show results for the solutions $x_s(t)$, $p_s(t)$, $y_s(t)$, and $q_s(t)$.  The calculations here are for $\gamma=4$, but the qualitative shape of the curves is the same for all $\gamma$.  The mRNA and protein concentrations $x_s(t)$ and $y_s(t)$ are sigmoidal and the conjugate momenta $p_s(t)$ and $q_s(t)$ have single ``bumps", beginning and ending at $0$ for $t \to \pm \infty$.  The protein curves lag the mRNA ones, as can be expected. 

\begin{figure}[ht] 
	\begin{center}
		\includegraphics[keepaspectratio=true,scale=0.4]{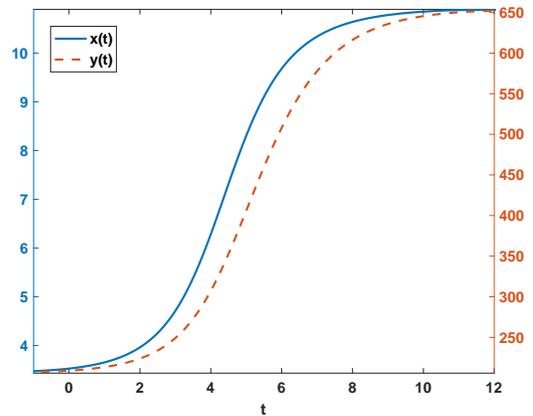}
		\caption{Typical concentration trajectories calculated using the relaxation method.  The blue (solid) line shows the mRNA concentration $x(t)$ and the red (dashed) line the protein concentration $y(t)$ .}				
	\label{figure5}
	\end{center}
\end{figure}

\begin{figure}[ht] 
	\begin{center}
		\includegraphics[keepaspectratio=true,scale=0.4]{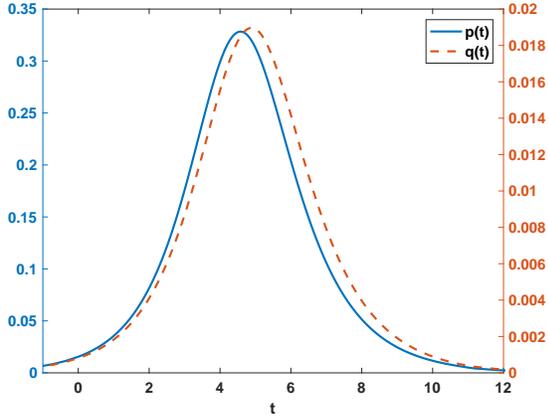}
		\caption{Typical momentum trajectories calculated using the relaxation method.  The blue (solid) line shows the mRNA concentration $x(t)$ and the red (dashed) line the protein concentration $y(t)$ .}
		\label{figure6}
	\end{center}
\end{figure}

We have calculated the extremal action $S_0$ for values of the mRNA degradation rate $\gamma$ from $1$ to $128$, using a fixed burst-size parameter $b=15$, so the translation rate $c = b\gamma$.   The values of the parameters in the Hill function (\ref{eq:Hillfn}) were taken to be those used for $b=15$ in the bursting-limit calculations described in subsection \ref{sect:burstsim}.   The calculations were done for both continuous- and discrete-protein-number models.  As explained above, in the limit $\gamma \to \infty$ we recover the bursting model of the preceding section. 

\begin{figure*}[ht] 
	\begin{center}
		\includegraphics[scale=0.73]{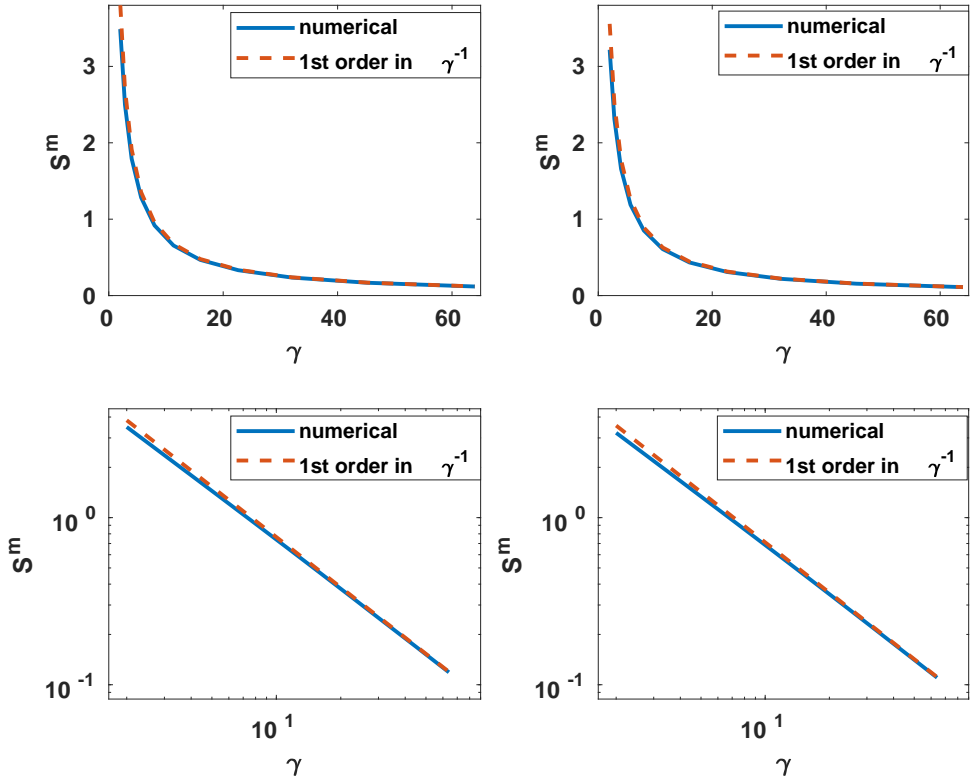}
		\caption{mRNA action as a function of $\gamma$ for continuous (left) and discrete (right) protein numbers. The lower plots are log-log, to make the $1/\gamma$ power-law behavior evident. Results of the numerical relaxation calculation are shown as blue (solid) lines and the first-order expansion in $1/\gamma$ as red (dashed) lines.	 }
		\label{figure7}
	\end{center}
\end{figure*}

\begin{figure*}[ht] 
	\begin{center}
		\includegraphics[keepaspectratio=true,scale=0.5]{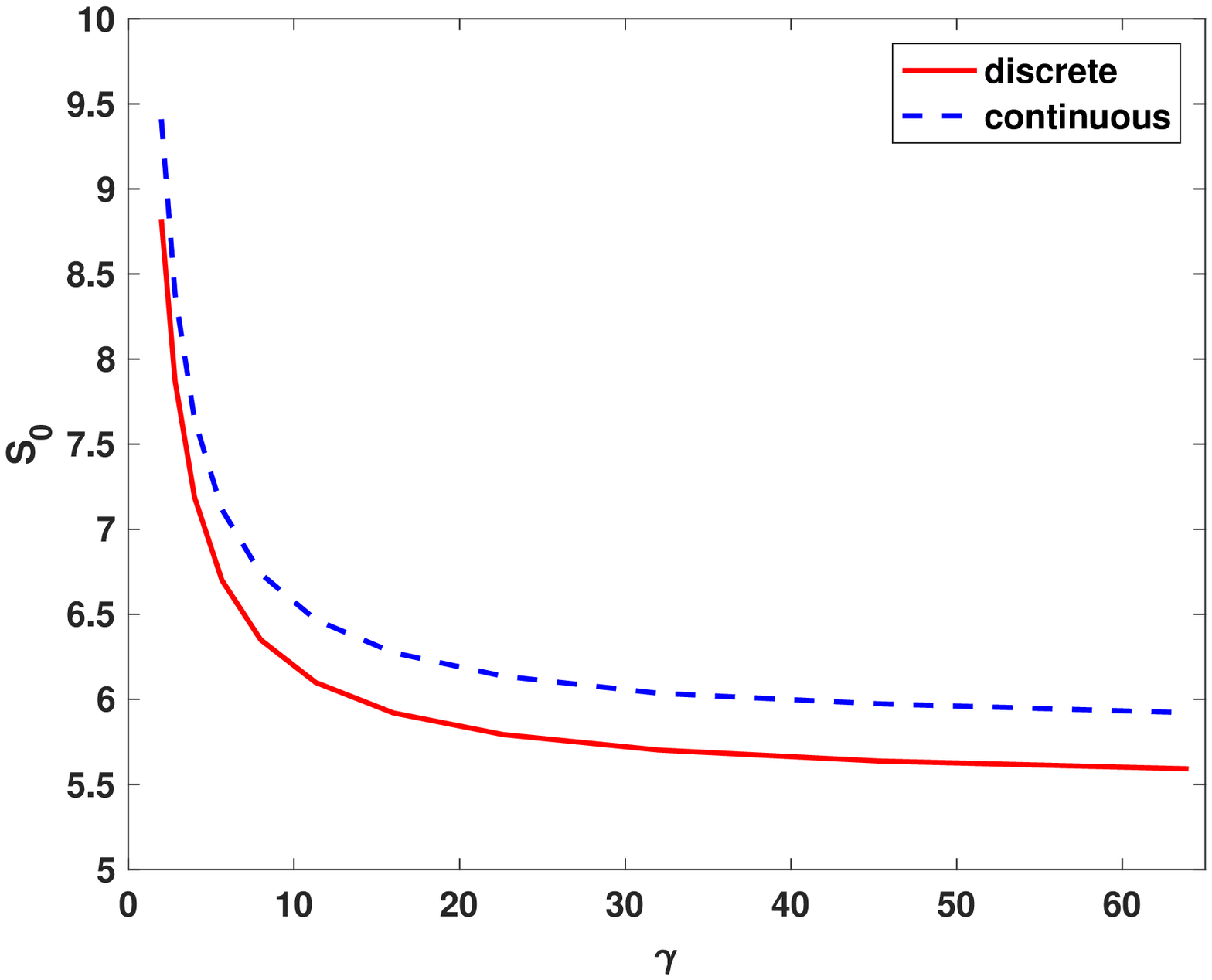}
		\caption{Total action as a function of $\gamma$. The solid (red) line shows the result for discrete protein number and the red (dashed) line that for continuous protein number.}
		\label{figure8}
	\end{center}
\end{figure*}

We examined the $\gamma$-dependence of the two terms in the extremal action (\ref{eq:S0full}), which we call the mRNA and protein actions, respectively.  These are shown in Figs. \ref{figure7} and \ref{figure8}.   The qualitative features are quite simple: The protein action is relatively insensitive to $\gamma$, and its large-$\gamma$ limit is consistent with the analytically calculated bursting-limit value.  The mRNA action, on the other hand, decreases rapidly with increasing $\gamma$.  At $\gamma=1$ the two actions are about the same size, while at $\gamma=128$ the mRNA term is two orders of magnitude smaller. The computed values are consistent with a $1/\gamma$ dependence over most of the range of $\gamma$ for which the calculations were done.  


The value of $b$ that we use is fairly large, but the calculations show a measurable difference between continuous and discrete protein number models.    For the protein action,  the discrete-case values are about $6\%$ smaller than the corresponding continuous-case ones (as found above in the bursting limit from eqns. (\ref{eq:Sol})-(\ref{eq:WSol}) and (\ref{eq:logq})-(\ref{eq:SyqHBd})), while for the mRNA action the  discrete-case values are about $10\%$ smaller. While these differences are not large, they occur in the exponent of the expression (\ref{eq:Gamma}) for the escape rate, so they can be important.  In the present case, the rate can be reduced by $30\%$ or so.

\subsection{Expanding around the bursting limit}
 
The above results show that the dependence of the extremal action (and thus, the escape rate from the metastable low-protein-number state) on the mRNA lifetime comes almost entirely from the mRNA term.  Its $\gamma$-dependence can be studied analytically for large $\gamma$ by expanding around the bursting limit.   

The calculation is simple in principle.  We start from the definition
\bea
S^m=\int_{x_{0}}^{x_{1}}p \, dx,\lb{eq:Sm}
\eea
where $x_{0,1} = y_{0,1}/c$, and use the bursting-limit conditions $\dot{x}=0$ and $\dot{p}=0$ and the fact that $H=0$ to write $x$ and $p$ in terms of $y$. We can then write
\bea
S^m=\int_{y_{0}}^{y_{1}}p(y) \frac{dx(y)}{dy}dy.\lb{eq:Smy}
\eea
The integrals can be evaluated in terms of elementary functions for both continuous and discrete protein numbers if the Hill function (\ref{eq:Hillfn}) has index $h=4$.  In both cases, we find $S^m \propto 1/\gamma$.  The calculations, which are straightforward but a bit messy, are relegated to Appendix B. 

For the continuous case, $S^m$ is simply proportional to $1/(b\gamma)$ (remembering we are always holding $bg(y)$ independent of $b$).  Since we have also observed that the protein action is almost independent of $\gamma$ for large $\gamma$ and we know that in this limit it is proportional to $1/b$,
this means that the total action has the form 
\be
S_0 = \frac{c_1 + c_2/\gamma}{b},
\ee
where $c_1$ and $c_2$ are constants evaluated in appendix B.

For the discrete case, $S^m$ turns out to have the form
\be
S^m = \frac{1}{\gamma}\left[ \frac{c_2}{b+1}+\frac{c_3}{(b+1)^2}\right],			\lb{eq:Smdisc}
\ee
with the constants evaluated in Appendix B.  Again, $S^m$ is proportional to $1/\gamma$, but with terms proportional to both $1/(b+1)$ and $1/(b+1)^2$.  In the large-$b$ limit,  (\ref{eq:Smdisc}) reduces to the continuous-case result. 

The straight solid lines in the log-log plots in Fig. \ref{figure7} are based on these results; one sees that they agree quite well with the results of the relaxation calculations all the way down to $\gamma = O(1)$. This is rather remarkable: In our derivations, we have used the conditions $\dot{x}=0$ and $\dot{p}=0$, which are only true in the bursting limit, so our results can only be correct to lowest order in $1/\gamma$ (i.e., the coefficient of $1/\gamma$ is evaluated at $\gamma=\infty$).  Nevertheless, as we see, this approximation is a good one over the entire range of $\gamma$ studied here.  Thus, the bursting limit is not just interesting in itself; it also allows us to treat the most important quantity in the general finite-$\gamma$ problem -- the action or activation barrier --  analytically.  

\subsection{Prefactors}

Like that of the action, the calculation of the prefactor for the full models is nontrivial and must be done numerically.   The classical calculation of Eyring \cite{eyring1935activated} applies only when the local flow in the rate equations can be written as the derivative of a potential function and the stochasticity is simple diffusion, with a concentration-independent diffusion constant.   A procedure for treating systems where the flow is not everywhere potential-derivable, based on a WKB scheme was set out by Maier and Stein \cite{MaierStein92}.  (Their treatment was still restricted, however, to systems with concentration-independent diffusion.)  Their scheme was extended to birth-death processes by Roma {\rm et al} \cite{roma2005optimal} for a model with mutual competition between two proteins and treated more generally by Bressloff \cite{bress2}.   The basic idea is to construct differential equations describing the evolution of the probability density over the course of the escape event, starting from the initial metastable distribution near $(x_0,y_0)$ and ending with the steady-state flow across the saddle point $(x_1,y_1)$, assuming that it remains Gaussian and centered on the trajectory. 
 
However, in a subsequent paper Maier and Stein \cite{maierstein1} showed that their earlier scheme applied only to systems where the local flow near the saddle point $(x_1,y_1)$ was derivable from a potential.  They found that, except for those exceptional systems, the final ($t \to \infty$) density near $(x_1,y_1)$ is not Gaussian because that point is not accessible from all directions in the local flow.   Our model is not in this special class.  (The model of Roma et al \cite{roma2005optimal}, on the other hand, is in the special class, so their calculation is valid.) Maier and Stein investigated the general case and showed, in an elaborate calculation, how one can find moments of the true exit distribution.  However, they did not go so far as to give an explicit result for the value of the prefactor, and, as far as we know, no such calculation has been done to date.

The Maier-Stein analysis does not tell us how bad an error one makes if one uses their earlier approach.  Therefore, here we perform the calculation this way and compare the result with simulations.

The derivation is analogous to that given in Appendix A.2 for the bursting model, though a bit more complicated because now there are two kinds of molecules (mRNA and protein).  It gives the formula
\be
\eta =\dfrac{\lambda_+}{2\pi}\sqrt{\dfrac{\det(Z_0)}{|\det(Z_1)|}}\dfrac{K_1}{K_0},	\lb{eq:Aeqn}
\ee
differing from the classical Eyring formula \cite{eyring1935activated} in the presence of the last factor.
Here,
\be
\lambda_+ = \frac{-(1+\gamma)+\sqrt{(1+\gamma)^2 + 4\gamma(bg'(y_1)-1)}}{2}
\ee
is the positive eigenvalue of the rate equation matrix at the unstable fixed point, $K_{0,1}$ are the limits as $t \to \pm \infty$ of the solutions of the differential equation 
\be
\frac{\d \ln K(t)}{\d t} = H_{xp} +H_{yq} + \half (Z_{xx}H_{pp} + Z_{yy}H_{qq}),   \lb{eq:Keqn}
\ee
and $Z_{0,1}$ are the corresponding limits of the symmetic $2 \times 2$ matrix function of $t$
\be 
Z =\begin{pmatrix}
Z_{xx} & Z_{xy} \\
Z_{yz} & Z_{yy}
\end{pmatrix}
= \begin{pmatrix}
\frac{\partial p}{\partial x} & \frac{\partial p}{\partial y} \\
\frac{\partial q}{\partial x} & \frac{\partial q}{\partial y}
\end{pmatrix},
\ee
which solves
\be
-\dot{Z} = ZBZ + ZA + A^{T}T + C.						\lb{eq:Zeqn}
\ee
The elements of the matrices $A$, $B$, and $C$ are second derivatives of $H$:
\be
A = \begin{pmatrix}
H_{px} & H_{py} \\
H_{qx} & H_{qy}
\end{pmatrix},
\ee
\be
B = \begin{pmatrix}
H_{pp} & H_{pq} \\
H_{qp} & H_{qq}
\end{pmatrix},
\ee
and
\be
C = \begin{pmatrix}
H_{xx} & H_{xy} \\
H_{yx} & H_{yy}
\end{pmatrix}.
\ee
These matrices are functions of $t$ through their dependence on the solutions $x$, $y$, $p$ and $q$ of the Hamilton equations of motion.  For the discrete-protein-number model, they are, explicitly, 
\be
A = \begin{pmatrix}
-\gamma \e^{-p} & cg'(cy)\e^{p} \\
\e^{q/c} & -\e^{-q/c} 
\end{pmatrix},
\ee
\be
B =  \begin{pmatrix}
g(cy)\e^p + \gamma x\e^{-p} & 0 \\
0 & (x \e^{q/c} + y \e^{-q/c})/c
\end{pmatrix},
\ee
and
\be
C =  \begin{pmatrix}
0 & 0 \\
0 & c^2 g'' (cy)(\e^p-1)
\end{pmatrix}.
\ee
For the continuous-protein model, they are different only in that $A_{21} = A_{22} = 1$ and $B_{22}=0$. (Here, and in the calculation described below, we have rescaled the protein concentration $y$ by a factor $1/c$, so that $y=x$ at the fixed points, and, correspondingly, rescaled $q$ by a factor $c$.)

\begin{figure*} 
	\begin{center}
		\includegraphics[height=7.5cm, width=18cm]{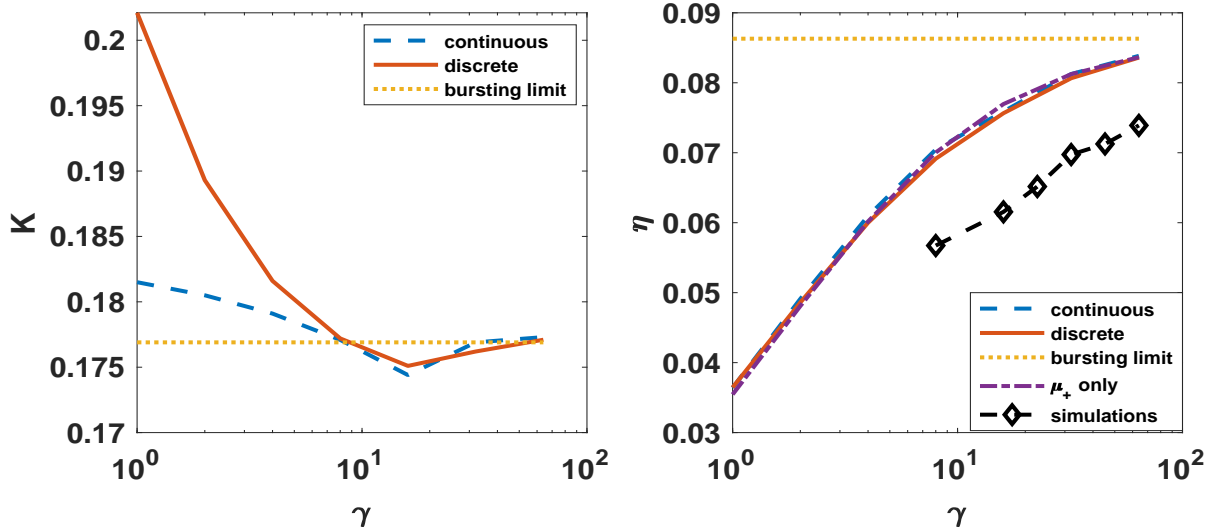}
		\caption{Left panel: The factor $K_1/K_0$ appearing in the prefactor formula (\ref{eq:Aeqn}) as a function of $\gamma$.  The blue (dashed) line shows the continuous protein number case and the red (solid) line the discrete protein-number case (burst size $b=15$). The yellow (dotted) line indicates the bursting-limit value $0.1769$ for this $b$.   Right panel: The prefactor $\eta$ as a function of $\gamma$, calculated from Eqn. (\ref{eq:Aeqn}) for continuous- and discrete-protein-number models ($b=15$).  Blue (dashed) and red (solid) lines indicate continuous and discrete protein number, as in the left-hand panel, and the purple (dot-dashed) line shows the result of assuming that $\eta$ depends on $\gamma$ only through the factor $\lambda_+$, i.e., that $\sqrt{\det(Z_0)/|\det(Z_1)|}\cdot K_1/K_0$ is constant (and equal to its value at $\gamma = \infty$, which can be obtained by requiring consistency with the known value of $\eta$ in the bursting-limit model).  The diamonds show the prefactor values inferred from simulations for $8 \le \gamma \le 64$.}
		\label{figure9}
	\end{center}
\end{figure*}

We have solved (\ref{eq:Zeqn}) numerically for $Z$ for both discrete and continuous-protein cases, using the same relaxation method employed above in finding the optimal path $x(t),y(t),p(t),q(t)$.  (As in previous calculations, we use a burst size $b=15$.)  Putting the elements $Z_{xx}(t)$ and $Z_{zz}(t)$ into (\ref{eq:Keqn}) and integrating, we evaluate $K_1/K_0$ and thus the prefactor $\eta$ from (\ref{eq:Aeqn}).  The results are shown in Fig. \ref{figure9}.  Evidently, the $K$'s for the two models differ slightly when $\gamma$ is not too large, but the prefactors are nearly the same for all $\gamma$.  Furthermore, the product $\sqrt{\det(Z_0)/|\det(Z_1)|}\cdot K_1/K_0$ is almost independent of $\gamma$; the $\gamma$-dependence of $\eta$ is accounted for entirely (within our numerical accuracy) by that of the rate equation eigenvalue $\lambda_+$.  We are tempted to conjecture that this is exact  (within the assumptions of the present approach), but we have not been able to prove it.  

We have simulated the discrete-protein-number model using the Gillespie algorithm, estimating the mean first passage time to reach the protein concentration $y_1$ by averaging over $10,000$ trials, for $b=15$ and $6$ values of $\gamma$ from $8$ to $64$.      We define an empirical prefactor by multiplying the  transition rate measured in simulations by $\exp(S_0)$, where $S_0$ is the barrier calculated in the preceding subsection. The empirical prefactors are marked with diamonds in the right panel of Fig. \ref{figure9}.  They show clearly that the naive theoretical calculation is wrong, as Maier and Stein would have anticipated.  The theoretical prefactors are around $15$\% higher than those found in the simulations.  We expect that as $\gamma \to \infty$ both the theoretical and the empirical values should approach the bursting-limit prefactor, so the discrepancy should disappear.

\section{Summary and Discussion}

We have presented a nearly-complete analysis of rare fluctuation-induced transitions between phenotypes in a minimal model of an autoregulatory genetic circuit.  At the most basic level of understanding -- that of the activation barrier or ``action'', much can be done analytically, using the bursting limit as a starting point in the path-integral formalism. In that limit, the action for the continuous-protein-number model can be evaluated quite simply (Eqns. (\ref{eq:Sol}) and (\ref{eq:WSol})).  For the full problem, we find (from numerical calculations) that the protein part of the action depends only weakly on the degradation rate ratio $\gamma$, so it can be approximated quite well by its bursting-limit value.  Furthermore, the mRNA part, which we find numerically to be quite accurately proportional to $1/\gamma$, can be calculated analytically.   Thus, to quite a good approximation, we can calculate everything about the action analytically for this model.  And it is only a little harder for the discrete-protein-number model  -- the only thing that we cannot calculate analytically is the (protein) action in the bursting limit.  However, it is simple to calculate it by expanding to a few orders in $1/b$, which should suffice for most interesting values of $b$.

In principle, the calculation of prefactors ought to be doable by expanding the exponent in the path integral to second order in the fluctuations and performing the resulting Gaussian functional integral.  We have not been able to do the calculation that way and so have resorted instead to solving Chapman-Kolmogorov (for the continuous-protein case) or solving Master equations using a WKB {\em Ansatz} (for discrete protein number).   Nevertheless, these approaches have yielded exact analytic expressions for the prefactors in the bursting limit (and the result does not depend on whether the protein number is continuous or discrete).

For the full problem (finite $\gamma$), guided by the work of Maier and Stein \cite{MaierStein92} and others \cite{roma2005optimal,bress2}, we performed a numerical WKB calculation of the prefactor.  The results are nearly, though not exactly, the same for discrete and continuous protein number (at least for the admitted somewhat large value of the parameter $b$ that we have used).   However, although they are of the right order of magnitude and their dependence on $\gamma$ seems qualitatively correct, their values are not in quantitative agreement with the simulation results -- they are 15-20\% too large.  Such a result could be anticipated from the work of Maier and Stein \cite{maierstein1}, but, to our knowledge, our result is the first measurement of the size of the error due to the faulty implicit assumptions of the naive theory.  Because the Maier-Stein result is generic, we think it would be important both to carry out a correct calculation and to study how the error due to the naive theory varies across a variety of systems.

We have studied a very simple model here in order to make the mathematics as transparent as possible. Real gene regulation networks are more complicated, but many such networks exhibit autoregulation and translational bursting.  For them, it would be interesting, if possible, to manipulate experimentally the burst size and the parameters describing the transcription rate (the minimal and maximal rates, the Hill index $h$ and dissociation constant $K$).  One could explore how the escape rate depends on them and to what extent these dependences are described by our simple bursting-limit model.  Away from the bursting limit, one could also test the $1/\gamma$ dependence on the mRNA degradation rate that we have found here.

As noted in the introduction, we believe that the methods employed here can be extended to more complex networks, such as the one involved in the {\em lac} operon in bacteria (see, for example, Sect. 6.4 of \cite{bress}).  There has been extensive modeling of this system at the level of rate equations (i.e., ignoring fluctuations) \cite{YildirimMackey,Yildirimetal}, and simulation studies with fluctuations \cite{Bhogale}.   Analytic studies at the level of ours here do not seem to have been done, but we think this is a feasible project. 

Among other phenomena where one might apply our methods, we name, in particular, cell differentiation.  This has been studied in simulations \cite{Chickarmane}, but not theoretically.  It would be especially interesting to calculate the rate of (rare) backward transitions, for example.

\vspace{1cm}
 
\appendix 

\section{Prefactors}
\subsection{Fokker-Planck equation with $x$-dependent diffusion}\lb{append:cont}

We follow here the approach of Dhar \footnote{https://home.icts.res.in/~abhi/notes/kram.pdf}.  Consider the Fokker-Planck equation
\bea \lb{eq:FPe}
\partial_tP(x,t)&=& -\partial_x[v(x)P(x)]+\half\partial_x^2 [D(x)P(x)] \nonumber \\ &=&-\partial_xJ(x),
\eea
obtained from the Ito interpretation of the stochastic differential equation
\be
dx = v(x)dt + \sqrt{D(x)}dW.
\ee
We consider the case where the system is bistable, with a metastable state at $x_0$, an unstable state at $x_1>x_0$ and a much more probable locally stable state at $x_2>x_1$. 
We will calculate the rate of escape from the neighbourhood of $x_0$ over the barrier around $x_1$, assuming a negligible rate for the backward transition from $x_2$ to $x_0$. 
In such a case, there is a small current $J_0$, independent of $x$, and 
\bea 
J_0=v(x)P(x)-\half \partial_x\big (D(x)P(x)\big ). \lb{eq:J0}
\eea
It is convenient to define $q(x) = D(x) P(x)$.  In equilbrium, $J_0=0$ and (\ref{eq:J0}) has the solution 
$q(x) = \exp [-W(x)]$, where $W'(x) = -2v(x)/D(x)$.  For $J_0 \neq 0$, consider the quantity 
$\partial_x\big (e^{W(x)}q(x)\big )$. Then $\partial_x\big (e^{W(x)}q(x)\big )=-2J_0e^{W(x)}$.  
Integrating this equation, we get
\bea
&&\int_{x_0}^{x_2}\partial_x\big (e^{W(x)}q(x)\big )dx = \nonumber \\ &=&q(x_2)e^{W(x_2)}-q(x_0)e^{W(x_0)} \\ &=&-2J_0\int_{x_0}^{x_2}e^{W(x)}dx. \nonumber
\eea
We are in a state where almost all of the probability is in the metastable region around $x_0$, so the first term on the left-hand-side is negligible, so we have
\bea
q(x_0)e^{W(x_0)}=2J_0\int_{x_0}^{x_2}e^{W(x)}dx.
\eea
Furthermore, $J_0=p_0\Gamma$, where $p_0$ is the probability to be in the metastable region and $\Gamma$ is the escape rate. Thus, the escape rate is 
\bea \lb{eq:r}
\Gamma =\frac{q(x_0)e^{W(x_0)}}{2p_0\int_{x_0}^{x_2}e^{W(x)}dx}.
\eea
Using Laplace's approximation on the integral in the denominator of equation (\ref{eq:r}), we have
\bea
\Gamma=\frac{q(x_0)e^{W(x_0)}}{2p_0e^{W(x_1)}\sqrt{2\pi / |W^{''}(x_1)|}}, 
\eea
and evaluating $p_0\approx (q(x_0)/D(x_0))\sqrt{2\pi/W^{''}(x_1)}$ we finally get 
\bea \lb{eq:res}
&\Gamma= \\ =& {\displaystyle \frac{1}{2\pi}\sqrt{\frac{D(x_0)}{D(x_1)}}\sqrt{|v^{'}(x_0)|v^{'}(x_1)}e^{W(x_0)-W(x_1)}.} \nonumber
\eea

We remark that if we had been using the Stratonovich convention for interpreting the underlying stochastic differential equation instead of the Ito one, the Fokker-Planck equation would have the form
\bea \lb{eq:FPe2}
\partial_tP(x,t)=&-\partial_x[v(x)P(x)] \\ &+\half\partial_x\big[ \sqrt{D(x)}\partial_x \big(\sqrt{D(x)}P(x)\big)\big] \nonumber 
\eea
instead of (\ref{eq:FPe}).  Carrying through the analogous calcuations for this case, one finds that the prefactor is that same as if the noise were additive: the factor $\sqrt{D(x_0)/D(x_1)}$ in (\ref{eq:res}) is missing.

\subsection{WKB and matching asymptotics for discrete protein number}\lb{append:ac}

In this subsection we calculate the prefactor $\eta$ in the escape rate (\ref{eq:Gamma}) for the discrete-protein-number bursting limit model. This case requires different methods from those in the preceding section, as our stochastic variable is no longer continuous. The method we are using was developed in \cite{escuder} and later used in \cite{bress2}. 

From equation (\ref{eq:sde}) with the probability of jumps given by (\ref{eq:Pofw_d}) we can derive the master equation 
\be
\dfrac{\partial P(n,t)}{\partial t}=\sum_{k=-\infty}^{\infty}\left\lbrace \omega_{n-k,k}P(k,t)-\omega_{k-n,n}P(n,t) \right\rbrace      \lb{eq:Masterappen}
\ee
,where
\be
\omega_{k,n}= \left\lbrace
\begin{array}{lr}
0 & k\leq -2 \\
n & k=-1 \\
g(n) (1-r)r^k & k\geq 1
\end{array}
\right.
\vspace{2cm}
\ee
are the transition rates.  

We know, because of the shape of $g(n)$, that the solution to this master equation will converge to a double-well shape stationary solution, with sharp maxima around the states that solve the equation $bg(n)=n$.

 Actually, we know that $y_0,y_1$, and $y_2$ solve $bg(y)=y$ and that, for the cases we study, they are of order  $10^2$ to $10^3$. Therefore, we introduce a large parameter $N$ of that order and a reduced concentration variable $x=n/N$ of $O(1)$. 
Rewriting our equation (\ref{eq:Masterappen}) in the new variable $x=n/N$, we have

\begin{widetext}
\be
\dfrac{\partial \Pi}{\partial t}(x,t)=\sum_{j=-\infty}^{\infty}N\left\lbrace \Omega_{-j}\left( x+\frac{j}{N}\right)\Pi \left( x+\frac{j}{N}\right)  -\Omega_{j}\left( x\right)\Pi \left( x \right)\right\rbrace , \lb{eq:continulimt}
\ee
with $\Pi(x,t)=P(xN,t)$ and
\be
\Omega_k(x)=\left\lbrace
\begin{array}{lr}
0 & k\leq -2 \\
x & k=-1 \\
N^{-1}g(xN) (1-r)r^k & k\geq 1.
\end{array}
\right.  
\ee
\end{widetext}

We use a well known technique called asymptotic matching that works in the following way:
\begin{itemize}
\item First, we approximate the quasi stationary distribution near the unstable fixed point $x_1$ by solving the Fokker-Planck equation in this region, assuming a constant flux $J>0$.
\item Second, we use the WKB method between $x_0$ and $x_1$ in the region where the Fokker-Planck approximation is not valid.
\item We then match these two solutions in the region where both are valid, enabling us to obtain a formula for the rate of escape.
\end{itemize}
 
The smallest eigenvalue of the master equation is zero, and the next-smallest is exponentially small with respect to $N$. This second eigenvalue is the sum of the rates of the rare transitions from $x_0$ to $x_2$ and back.  Under our present assumptions, the backward rate is exponentially weaker than the forward one, so this sum of rates will be basically the rate of escape from $x_0$.

We define the  quasi-stationary probability distribution $\Pi(x)$ as the eigenfunction that corresponds to the second eigenvalue $r_-$, i.e. the rate of escape. The function $\Pi(x) e^{-r_{-}t}$ solves the Fokker-Planck approximation of (\ref{eq:Masterappen}) around $x_1$. We assume that the constant flux $J$ through $x_1$  is re-injected into the mestastable well around $x_0$, so $\Pi(x)$ remains stationary.
We then expand
\begin{widetext}
\bea \lb{eq:quastat1}
0&=&\sum_{j=-\infty}^{\infty}N\left\lbrace \Omega_{-j}\left( x+\frac{j}{N}\right)\Pi \left( x+\frac{j}{N}\right)-\Omega_{j}\left( x\right)\Pi \left( x \right)\right\rbrace \\ &=&\sum_{j=-\infty}^{\infty}N\left\lbrace \Omega_{j}\left( x-\frac{j}{N}\right)\Pi \left( x-\frac{j}{N}\right)-\Omega_{j}\left( x\right)\Pi \left( x \right)\right\rbrace , 
\eea
to second order in $1/N$:
\be
0=\sum_{k=-\infty}^{+\infty} N\left\lbrace \left( \Omega_j(x)-\dfrac{j}{N}\Omega_j'(x)+\dfrac{j^2}{2N^2}\Omega_j''(x)\right) \left( \Pi(x)-\dfrac{j}{N}\Pi'(x)+\dfrac{j^2}{2N^2}\Pi''(x)\right)-\Pi(x) \Omega_j(x)\right\rbrace. \lb{eq:taylexpan1}
\ee
We collect the terms proportional to powers of $1/N$:
\be
0=-\sum_{k=-\infty}^{\infty} j\left\lbrace \Omega_j'(x)\Pi(x)+\Omega_j(x)\Pi'(x)\right\rbrace
+\dfrac{1}{2N}\sum_{k=-\infty}^{\infty} j^2\left\lbrace \Omega_j''(x)\Pi(x)+ 2\Omega_j'(x)\Pi'(x)+\Omega_j(x)\Pi''(x)\right\rbrace . \lb{eq:taylexpan2}
\ee
\end{widetext}

This has the form of a stationary Fokker-Planck equation

\be
0=\dfrac{\partial}{\partial x}\left\lbrace A(x)\Pi(x)\right\rbrace -\dfrac{1}{2N} \dfrac{\partial^2}{\partial x^2} \left\lbrace B(x)\Pi(x)\right\rbrace
\ee

with
\begin{align}
A(x)=\sum_{j=-\infty}^{\infty}j\Omega_j = bN^{-1}g(Nx)-x \\
B(x)=\sum_{j=-\infty}^{\infty}j^2\Omega_j = (1+2b)bN^{-1}g(Nx)+x .
\end{align}
The flux is
\be
J=A(x)\Pi(x)-\frac{1}{2N}\dfrac{\partial}{\partial x}(B(x)\Pi(x)).
\ee
This can be solved for $\Pi$:

\bea
\Pi(x)&=&{\displaystyle \dfrac{2JN}{B(y)}e^{2N\int_{x_1}^{x} \frac{A(z)}{B(z)}dz}\int_{x}^{\infty}e^{-2N\int_{x_1}^{z} \frac{A(\theta)}{B(\theta)}d\theta}dz} \nonumber \\ &\simeq & \dfrac{JN}{x_1 (1+b)}e^{\frac{(x-x_1)^2}{2\sigma^2}}\int_{x}^{\infty} e^{-\frac{(z-x_1)^2}{2\sigma^2}}dz,
\eea

with 
\be
\sigma^2=\frac{x_1 (1+b)}{N(bg'(Nx_1)-1)} .\lb{eq:sigma}.
\ee
 We can simplify $\Pi$ in the regime $x \ll x_1-\sigma$
\be
\Pi(x)=
\dfrac{JN\sigma \sqrt{2\pi}}{x_1 (1+b)}e^{\frac{(x-x_1)^2}{2\sigma^2}}. \lb{eq:formatch}
\ee

Now we use the WKB method to obtain a solution valid in the range  $x_0 <x<x_1-\sigma$ that we will match with (\ref{eq:formatch}) in the region $ x \ll x_1-\sigma$, which will enable us to obtain $J$. 
 The WKB {\em ansatz} is $\Pi=K(x) e^{-N\mathcal{W}(x)}$.  With it we can expand:
\begin{widetext}
\be
\Pi \left( x\pm \frac{j}{N}\right)\simeq K\left( x\pm \frac{j}{N}\right)e^{-N \mathcal{W}\left( x\pm \frac{j}{N}\right) } \simeq \left\lbrace K(x) \pm \frac{j}{N}K'(x)
\right\rbrace  e^{-N \left[ \mathcal{W}(x)\pm \frac{j}{N}\mathcal{W}'(x)+\frac{j^2}{2N^2}\mathcal{W}''(x)\right] } .
\ee
Now, expanding $e^{-N^{-1}\mathcal{W}''}$ in $N^{-1}$, we obtain, consistently to order $1/N$,
\be
\Pi\left( x\pm \frac{j}{N}\right)\simeq \left\lbrace K(x) \pm \frac{j}{N}K'(x) \right\rbrace e^{-N\mathcal{W}(x)\mp j\mathcal{W}'(x) } \left\lbrace 1-\frac{j^2}{2N}\mathcal{W}''(x)\right\rbrace .
\ee
Also expanding $\Omega$ and using equation (\ref{eq:quastat1}) we obtain 
\be
0=\sum_{k=-\infty}^{\infty}\left\lbrace \left( \Omega_j-\dfrac{j}{N}\Omega'\right)\left( K-\dfrac{j}{N}K'\right)\left( 1-\dfrac{j^2}{2N}\mathcal{W}''\right)e^{-N\mathcal{W}+j\mathcal{W}'}-\Omega_j Ke^{-N\mathcal{W}}\right\rbrace . \lb{eq:taylorwkb1}
\ee
From the terms of $O(1)$ we have
\be
0=\sum_{j=-\infty}^{\infty}\Omega_j(x)\left( e^{j\mathcal{W}'(x)}-1\right),
\ee
which can be interpreted as a stationary Hamilton-Jacobi equation for the Hamiltonian 
\be
H(x,q)=\sum_{j=-\infty}^{\infty}\Omega_j(x)\left( e^{jq}-1\right)=(e^q-1)\left( \frac{bN^{-1}g(Nx)}{1-b(e^q-1)} - xe^{-q}\right),				
\ee
which coincides with (\ref{eq:HsubBd}). The $O(N^{-1})$ terms yield
\be
0=\dfrac{K'(x)}{K(x)}\sum_{j=-\infty}^{\infty}\left\lbrace j\Omega_j e^{j\mathcal{W}'}\right\rbrace +\sum_{j=-\infty}^{\infty}\left\lbrace j\Omega'_j e^{j\mathcal{W}'}\right\rbrace +\sum_{j=-\infty}^{\infty}\dfrac{j^2}{2}\left\lbrace \Omega_j \mathcal{W}''e^{j\mathcal{W}'}\right\rbrace .
\ee
This differential equation for $K$ can be written in terms of derivatives of the Hamiltonian as
\be
\dfrac{K'(x)}{K(x)}H_q(x,q(x))=-\dfrac{1}{2}q'(x)H_{qq}(x,q(x))-H_{qx}(x,q(x)) \lb{eq:prefactor1}
\ee
with $q(x)=\mathcal{W}'(x)$, which solves $H(x,q(x))=0$.
Thus, our WKB solution has the form
\be
\Pi(y)=K(x)e^{-N\mathcal{W}(x)} ,
\ee
\end{widetext}
where the action $\mathcal{W}$ is defined as $\mathcal{W}=\int_{x_0}^{x_1} q(x)dx$ and $K(x)$ solves(\ref{eq:prefactor1}). 

To obtain $K$ we just have to use the fact that $H(x,q(x))=0$ and take derivatives: 
\be
H_q q'+H_x=0 \lb{eq:solvingpre1}
\ee
and
\be
H_{qq}(q')^2+2H_{qx}q'+H_q q''+H_{xx}=0 ,\lb{eq:solvingpre2}
\ee
from which
\bea \lb{eq:solvingpre3}
-q'(x)H_{qq}(x,q(x))-2H_{qx}(x,q(x)) \nonumber \\=H_q \left( -\dfrac{H_{xx}}{H_x}+\dfrac{q''}{q'}\right) .
\eea
Thus, the equation for the prefactor can be rewritten
\be
2\dfrac{K'(x)}{K(x)}=\left( -\dfrac{H_{xx}}{H_x}+\dfrac{q''}{q'}\right).  \lb{eq:generalsol}
\ee
The solution of this equation is
\be
K(x)=\dfrac{C}{\sqrt{x(N^2x+g(Nx))}}. \lb{eq:activwkb}
\ee
We also know that the flux is the escape rate times the population in the metastable well:
\be
J = \Gamma \int_{-\infty}^{x_1}\Pi(x)dx. 			\lb{eq:fokflux}
\ee 
Using equation (\ref{eq:fokflux}) and the WKB solution, we can employ the Laplace method to obtain 
\be
\Gamma=\dfrac{J}{K(x_0)}\sqrt{\dfrac{N \mathcal{W}''(x_0)}{2\pi}}e^{N\mathcal{W}(x_0)}. \lb{eq:ratesc}
\ee
Approximating $K(x)\exp[-N \mathcal{W}(x)]$ by a second order Taylor expansion around $x_1$ and matching with (\ref{eq:formatch}), we obtain 
\begin{align}
\sigma^2=\dfrac{1}{N\left\vert \mathcal{W}''(x_1)\right\vert} \\
J=\dfrac{K(x_1)x_1(1+b)e^{-N\mathcal{W}(x_1)}}{\sigma \sqrt{2\pi}}.
\end{align}
Substituting this flux in equation (\ref{eq:ratesc}) yields the rate 
\begin{widetext}
\be
\Gamma=\dfrac{K(x_1)}{2\pi K(x_0)}x_1(1+b)\sqrt{\left\vert \mathcal{W}''(x_0)\mathcal{W}''(x_1)\right\vert}e^{-N(\mathcal{W}(x_1)-\mathcal{W}(x_0))}. \lb{eq:wkbrate1}
\ee
\end{widetext}
From (\ref{eq:activwkb}), it can be seen that $K(x_1)/K(x_0)=x_0/x_1=y_0/y_1$. Also, because $\mathcal{W}''=q'$,
\be
\mathcal{W}''(x_{0,1})=\dfrac{1}{1+b}\dfrac{1-bg'(Nx_{0,1})}{x_{0,1}},
\ee
so we can rewrite (\ref{eq:wkbrate1}) as
\be
\Gamma=\dfrac{1}{2\pi}\sqrt{\dfrac{y_0}{y_1}}\sqrt{( 1-bg'(y_0)) \left\lvert 1-bg'(y_1)\right\rvert}e^{-N\Delta \mathcal{W}}, \lb{eq:wkbrate2}
\ee
in agreement with (\ref{eq:prefac11}) with $N\mathcal{W}=S$.

\section{Expansion of the mRNA action in $1/\gamma$}
Here we expand the mRNA action $S^m$ around the bursting limit for both continuous and discrete protein number. 

\subsection{continuous protein number} \lb{append:conti}
We would like to evaluate the mRNA action
\bea
S^m=\int_{x_{0}}^{x_{1}}pdx,\lb{eq:App_S}
\eea
where $x_{0,1} = y_{0,1}/c$ and $x(t)$ and $p(t)$ are respectively the optimal path and conjugate momentum variable for the model with continuous number of proteins, with Hamiltonian given by equation (25). Then we can use our results from equations (\ref{eq:epofq}), (\ref{eq:xofyandq}) and (\ref{eq:qdot0}) to express $p$ and $x$ as functions of $q$ and $y$, namely
\bea
p=\ln\frac{y}{bg(y)} \quad  {\rm and}  \quad  x=\frac{1}{b\gamma} \frac{y^2}{b g(y)}.\lb{eq:App_px}
\eea
Substituting (\ref{eq:App_px}) into (\ref{eq:App_S}) leads to
\begin{widetext}
\bea
S^m=\int_{y_{0}}^{y_{1}}p(y)\frac{dx(y)}{dy}dy=\frac{1}{b\gamma}\int_{y_{0}}^{y_{1}}\ln\Big (\frac{y}{bg(y)}\Big )\frac{\partial}{\partial y }\Big (\frac{y^2}{bg(y)}\Big )dy,
\eea
and integrating by parts we get 
\bea
S^m=\Big [\frac{1}{b\gamma}\ln\Big (\frac{y}{bg(y)}\Big )\frac{y^2}{bg(y)}\Big ]_{y_0}^{y_1}-\frac{1}{b\gamma}\int_{y_{0}}^{y_{1}}y\frac{\partial}{\partial y}\Big (\frac{y}{bg(y)}\Big )=-\frac{1}{b\gamma}\int_{y_{0}}^{y_{1}}y\frac{\partial}{\partial y}\Big (\frac{y}{bg(y)}\Big )dy.
\eea 
The boundary term in the equation above vanishes because $y=bg(y)$ at the fixed points.
Integrating by parts again yields
\bea
S^m=\Big [-\frac{1}{b\gamma}\frac{y^2}{bg(y)}\Big ]_{y_0}^{y_1}+\frac{1}{b\gamma}\int_{y_0}^{y_1}\Big (\frac{y}{bg(y)}\Big )dy=-\frac{1}{b\gamma}(y_1-y_0)+\frac{1}{b\gamma}\int_{y_0}^{y_1}\Big (\frac{y}{bg(y)}\Big )dy
\eea
The production rate $g(y)$ is a Hill function given by equation (\ref{eq:Hillfn}). If we put it into the above equation we get
\bea
S^m=-\frac{1}{b\gamma}(y_1-y_0)+\frac{1}{2b\gamma\alpha}(y_1^2-y_0^2)-\frac{1}{b\gamma}\Big (\frac{\beta}{\alpha^2}\Big )\int_{y_0}^{y_1}\frac{y^{h+1}}{K^h+(1+\beta/\alpha)y^h}dy,\lb{eq:nastan}
\eea
where $\alpha=ab$ and $\beta=bg_0$.
For $h=4$ we can evaluate analytically the integral in equation (\ref{eq:nastan}) analytically, which leads to
\bea
S^m=-\frac{1}{b\gamma}\Bigg\{ (y_1-y_0)-\frac{1}{b}\Bigg [\frac{1}{2(a+g_0)}(y_1^2-y_0^2)  \\
+\frac{K^2g_0}{2\sqrt a(a+g_0)^{3/2}}\Bigg ( \tan^{-1}\Big (\frac{y_1^2}{K^2}\sqrt{1+\frac{g_0}{a}}\Big)- \tan^{-1}\Big (\frac{y_0^2}{K^2}\sqrt{1+\frac{g_0}{a}}\Big)\Bigg )\Bigg ]\Bigg\}
\eea
\end{widetext}

\subsection{discrete protein number} \lb{append:disc}

For the model with discrete protein number, with Hamiltonian given by equation (\ref{eq:hamilt}), we start from equations (\ref{eq:gam3}), (\ref{eq:epofq2}) and (\ref{eq:logq}):
\be
p(y) = \ln \left( \frac{y+g(y)}{(1+b)g(y)} \right) \lb{eq:App_p}
\ee
and
\be
x(y) = \frac{1}{\gamma (1+b)^2}  \frac{(y + g(y))^2}{g(y)} . \lb{eq:App_x}
\ee
Substituting (\ref{eq:App_p}) and (\ref{eq:App_x}) into (\ref{eq:App_S}) and integrating by parts we get
\begin{widetext}
\bea
S^m &=& -\frac{1}{\gamma (1+b)^2} \int_{y_0}^{y_1} \d y
\frac{(y+g(y))^2}{g(y)} \cdot \frac{(1+b)g(y)}{y+g(y)} \cdot \frac{\d}{\d y}
\left( \frac{y+g(y)}{(1+b)g(y)} \right) 					\nonumber \\ 
&=&  -\frac{1}{\gamma (1+b)^2} \int_{y_0}^{y_1} \d y (y+g(y)) \frac{\d}{\d y} \left( \frac{y+g(y)}{g(y)} \right).
											\lb{eq:ti}
\eea
Now, writing $bg(y) = \beta(y)$ (remember $\beta$ is independent of $b$), we have
\be
S^m = -\frac{b}{\gamma (1+b)^2} \int_{y_0}^{y_1} \d y \left(y+\frac{1}{b}\beta(y)\right) 
\frac{\d}{\d y} \left( \frac{y}{\beta(y)} \right). 
\ee
Integrating by parts again yields the result
\bea
S^m &=& -\frac{1}{\gamma (1+b)^2} \left[ (1+b)(y_1-y_0) 
- b \int_{y_0}^{y_1} \d y  \frac{y}{\beta(y)}
- \int_{y_0}^{y_1} \d y\, y \frac{\beta'(y)}{\beta(y)}  \right] 			  		\nonumber \\
&=& \frac{1}{\gamma (1+b)^2} \left[ (1+b)(y_0-y_1) +b I_1 + I_2 \right],  \nonumber \\
&=& \frac{1}{\gamma} \left[ \frac{y_0-y_1+I_1}{b+1} +\frac{I_2-I_1}{(b+1)^2}\right],	\lb{eq:genres}
\eea
with
\be
I_1 = \int_{y_0}^{y_1} \d y \frac{y}{\beta(y)}
\ee
and
\bea
I_2 &=& \int_{y_0}^{y_1} \d y \frac{y\beta'(y)}{\beta(y)} = \int_{y_0}^{y_1} \d y \, y \frac{\d}{\d y} \ln \beta(y)=	 y_1 \ln \beta(y_1) - y_0 \ln \beta(y_0) - \int_{y_0}^{y_1} \d y \ln \beta(y)=  \nonumber \\
&=& y_1 \ln y_1 - y_0 \ln y_0 - \int_{y_0}^{y_1} \d y \ln \beta(y)    \lb{eq:formofI2}
\eea
$I_1$ and $I_2$ are independent of $b$; thus, the dependence of $S^m$ on $\gamma$ and $b$ is quite simple.

The integrals in $I_1$ and $I_2$ can be done analytically if we take $g(y)$ to be the Hill function (\ref{eq:Hillfn}) with $h=4$. Explicitly, we have, for $I_1$,
\bea
I_1 &=& \int_{y_0}^{y_1} \d y \frac{y}{ab+ \frac{bg_0y^4}{y^4 + K^4}} = \int_{y_0}^{y_1} \left( \frac{y^4 + K^4}{(a+g_0)by^4 + abK^4} \right) y\d y,
\eea
and by making the change of variable $u=y^2$ we get
\bea
I_1 &=& \half \int_{y_0^2}^{y_1^2} \d u \left( \frac{u^2 + K^4}{(a+g_0)bu^2 + abK^4} \right) = \frac{1}{2ab} \int_{y_0^2}^{y_1^2} \d u \left( 1- \frac{bg_0u^2}{(a+g_0)bu^2 + abK^4} \right) \nonumber \\
&=& \frac{y_1^2-y_0^2}{2ab} - 
\frac{g_0}{2ab(a+g_0)} \int_{y_0^2}^{y_1^2} \d u \frac{u^2}{(u^2 + (a/(a+g_0))K^4}		 \\
&=& \frac{y_1^2-y_0^2}{2ab} -  \frac{g_0}{2ab(a+g_0)}\int_{y_0^2}^{y_1^2} \d u 
\left[ 1 - \left(\frac{aK^4}{a+g_0}\right) \frac{1}{u^2 + (a/(a+g_0))K^4} \right]		\nonumber \\
&=& \frac{y_1^2-y_0^2}{2(a+g_0)b} + \frac{g_0K^4}{2(a+g_0)^2b}
\int_{y_0^2}^{y_1^2} \frac{\d u} {u^2 + (a/(a+g_0))K^4}					\nonumber \\
&=& \frac{y_1^2-y_0^2}{2(a+g_0)b} + \frac{g_0K^2}{2a^{1/2}(a+g_0)^{3/2}b} \left[
\tan^{-1} \left( \frac{y_1^2}{K^2}\sqrt{\frac{a+g_0}{a}}\right) 
- \tan^{-1} \left( \frac{y_0^2}{K^2}\sqrt{\frac{a+g_0}{a}}\right) \right].\nonumber
\eea
 
For $I_2$, the integral in the third term of (\ref{eq:formofI2}) is
\bea
 \int_{y_0}^{y_1} \d y \ln \beta(y) &=&  \int_{y_0}^{y_1} \d y \ln 
 \left( ab + \frac{bg_0 y^4}{y^4 + K^4} \right) = \int_{y_0}^{y_1} \d y \ln \left( \frac{(a +g_0)b y^4 + abK^4}{y^4 + K^4} \right)    \nonumber \\
 &=&  (y_1-y_0) \ln[(a+g_0)b] + \int_{y_0}^{y_1} \d y \ln \left( \frac{y^4 + (a/(a+g_0))K^4}{y^4 + K^4} \right) .
 \eea
In this expression we have the difference of two integrals of the form $\int_{y_0}^{y_1} \d y \ln (y^4 + c)$ with  $c = aK^4/(a+g_0)$ in the first integral, and $c = K^4$ in the second one. The indefinite integral is 
\bea
\int \d y \ln (y^4 + c) &=& y[\ln(c+y^4) -4]+ \sqrt{2} c^{1/4} \left[ \tanh^{-1}\left(\frac{\sqrt{2} c^{1/4}y}{\sqrt{c} + y^2} \right) \right.							\nonumber \\
&-& \left. \tan^{-1}\left( 1 - \frac{\sqrt{2}y}{c^{1/4}} \right) 
            + \tan^{-1}\left(1 + \frac{\sqrt{2}y}{c^{1/4}} \right) 
\right] 
\eea
We denote this expression $F(y,c)$, so we can write
\bea
I_2 &=& y_1 \ln y_1 - y_0 \ln y_0 - (y_1-y_0)\ln[(a+g_0)b]  - F(y_1, aK^4/(a+g_0))   \nonumber \\
&+&  F(y_0, aK^4/(a+g_0)) +F(y_1, K^4) - F(y_0, K^4).
\eea
Thus, all terms in formula (\ref{eq:genres}) can be expressed in terms of elementary functions.

\pagebreak

\end{widetext}

\section{Relaxation method for numerical integration of Hamilton equations}\lb{append:relax}

The method is essentially a high-dimensional version of the Newton-Raphson method: If we want to find a solution $u_0$ of $f(u) =0$ starting from a guess $u_1$, one expands $f(u)$ around $u_1$
\be
f(u) = f(u_1) + (u-u_1)f'(u_1),
\ee
leading to a new guess, $u_2 = u_1 - f(u_1)/f'(u_1)$ for the root.  One then iterates this, taking
\be
u_{n+1} = u_n - (f'(u_n))^{-1}f(u_n)
\ee
until convergence is achieved.  In our case, the $u$'s are the values of $x_s(t_i)$, $p_s(t_i)$, $y_s(t_i)$, and $q_s(t_i)$ on a set of $N$ closely-spaced points $t_i$.  Boundary conditions specify 4 of these (for example, $x_s(t_1)$, $y_s(t_1)$, $x_s(t_N)$ and $y_s(t_N)$, so our $u$ is a $N$-$4$-dimensional vector. (Typically, we take $N=100.) $    Our counterpart of the function $f$ that should be equal to zero is a time-discretized version of the Hamilton equations for each time step from $t_i$ to $t_{i+1}$ (also $N$-$4$-dimensional).  Finally, the counterpart of the derivative $f'(u)$ is a $N$-$4 \times N$-$4$ matrix: the partial derivatives of the $N$-$4$ discretized Hamilton equations with respect to the $N$-$4$ free values of $x_s(t_i)$, $p_s(t_i)$, $y_s(t_i)$, and $q_s(t_i)$.  The multiplication $(f'(u_n))^{-1}fu_n)$ is then a matrix-times-vector operation.

A small detail:  we have found it helpful to fix the initial and final boundary values of $p_s(t)$ and $q_s(t)$ as well as those of $x_s(t)$ and $y_s(t)$.  Then the derivative matrix is no longer square: there are more conditions ($N$-$4$) than variables  ($N$-$8$): The problem is overdetermined, and all we can do is to choose the solution that minimizes an error measure.  Fortunately, the Matlab operation $\backslash$,  does this automatically:  If $\vec{x}$ is an $n$-component vector and $B$ is an $m \times n$ matrix with $m>n$, then $y = B\backslash x$ returns the $n$-component vector $\vec{y}$ that minimizes the quadratic error $(\vec{y}-B\vec{x})^2$. 

The functions $x_s(t)$, $p_s(t)$, $y_s(t)$, and $q_s(t)$ that we seek approach their fixed-point values $x_{0,1}$, $0$, $y_{0,1}$, $0$ exponentially in the limits $t \to \infty$.  We can find the exponents characterizing this approach and the relative magnitudes of the four variables by linearizing the Hamilton equations near the fixed points.  Consistency requires that we take the initial deviation from $(x_0,0,y_0,0)$ proportional to the unstable eigenvector for which all components have the same sign and the final approach to  $(x_1,0,y_1,0)$ proportional to the stable eigenvector where the signs of the $p$ and $q$ components are the same and opposite to those of the $x$ and $y$ components.  Our strategy then begins with assuming these exponential forms for $t<0$ and $t>T$, where the deviations from the fixed points at $0$ and $T$ small enough that we can trust these exponential approximations and $T$ is initially a guess about how long it takes the dynamics of the system to get from the neighborhood of one fixed point to the other.  We then use the relaxation method described above to find the values of $x_s(t_i)$, $p_s(t_i)$, $y_s(t_i)$, and $q_s(t_i)$ that solve the discretized Hamilton equations on the $t_i$ between $0$ and $T$, with the boundary values given by the assumed values there from the exponential tail approximation. In this way, one can hope to have a good approximation for the variables from $t=-\infty$ to $+\infty$.

Of course, the value of $T$ is initially just guessed.  Therefore, the procedure is repeated until the best $T$ is found.  There are several possible criteria for defining ``best''; we have considered three of them.  The first is simply the minimum square error obtained in the overdetermined relaxation process. Another indicator is the square errors in the derivatives of $x_s$, $p_s$, $y_s$ and $q_s$ at the boundaries $0$ and $T$ where the numerical relaxation solution is patched to the analytic exponential forms for $t<0$ and $t>T$.  The third criterion is based on the fact that if we had an exact calculation, $H(t)$ would vanish at all $t$, and so would $\int H(t)dt$.  In our approximate calculation, $H(t)$ is small but nonzero and so is its time integral.  We therefore seek for the value of $T$ for which $\int H(t)dt = 0$, i.e., the errors are unbiased.
Remarkably, in our computations these three criteria lead to nearly the same choices of optimal $T$, and the differences in the estimated values of $S_0$ are very small.



\bibliographystyle{apsrev4-1.bst}
\bibliography{activ}

\begin{thebibliography}{27}%
\makeatletter
\providecommand \@ifxundefined [1]{%
 \@ifx{#1\undefined}
}%
\providecommand \@ifnum [1]{%
 \ifnum #1\expandafter \@firstoftwo
 \else \expandafter \@secondoftwo
 \fi
}%
\providecommand \@ifx [1]{%
 \ifx #1\expandafter \@firstoftwo
 \else \expandafter \@secondoftwo
 \fi
}%
\providecommand \natexlab [1]{#1}%
\providecommand \enquote  [1]{``#1''}%
\providecommand \bibnamefont  [1]{#1}%
\providecommand \bibfnamefont [1]{#1}%
\providecommand \citenamefont [1]{#1}%
\providecommand \href@noop [0]{\@secondoftwo}%
\providecommand \href [0]{\begingroup \@sanitize@url \@href}%
\providecommand \@href[1]{\@@startlink{#1}\@@href}%
\providecommand \@@href[1]{\endgroup#1\@@endlink}%
\providecommand \@sanitize@url [0]{\catcode `\\12\catcode `\$12\catcode
  `\&12\catcode `\#12\catcode `\^12\catcode `\_12\catcode `\%12\relax}%
\providecommand \@@startlink[1]{}%
\providecommand \@@endlink[0]{}%
\providecommand \url  [0]{\begingroup\@sanitize@url \@url }%
\providecommand \@url [1]{\endgroup\@href {#1}{\urlprefix }}%
\providecommand \urlprefix  [0]{URL }%
\providecommand \Eprint [0]{\href }%
\providecommand \doibase [0]{http://dx.doi.org/}%
\providecommand \selectlanguage [0]{\@gobble}%
\providecommand \bibinfo  [0]{\@secondoftwo}%
\providecommand \bibfield  [0]{\@secondoftwo}%
\providecommand \translation [1]{[#1]}%
\providecommand \BibitemOpen [0]{}%
\providecommand \bibitemStop [0]{}%
\providecommand \bibitemNoStop [0]{.\EOS\space}%
\providecommand \EOS [0]{\spacefactor3000\relax}%
\providecommand \BibitemShut  [1]{\csname bibitem#1\endcsname}%
\let\auto@bib@innerbib\@empty
\bibitem [{\citenamefont {{T. B. Kepler and T. C. Elston}}(2001)}]{Kepler}%
  \BibitemOpen
  \bibfield  {author} {\bibinfo {author} {\bibnamefont {{T. B. Kepler and T. C.
  Elston}}},\ }\href@noop {} {\bibfield  {journal} {\bibinfo  {journal}
  {Biophys J}\ }\textbf {\bibinfo {volume} {81}},\ \bibinfo {pages} {3116}
  (\bibinfo {year} {2001})}\BibitemShut {NoStop}%
\bibitem [{\citenamefont {{P. C. Bressloff}}(2014)}]{bress}%
  \BibitemOpen
  \bibfield  {author} {\bibinfo {author} {\bibnamefont {{P. C. Bressloff}}},\
  }\href@noop {} {\emph {\bibinfo {title} {Stochastic Processes in Cell
  Biology}}}\ (\bibinfo  {publisher} {Springer},\ \bibinfo {year}
  {2014})\BibitemShut {NoStop}%
\bibitem [{\citenamefont {{P. C. Bressloff}}(2017)}]{bressjphysa}%
  \BibitemOpen
  \bibfield  {author} {\bibinfo {author} {\bibnamefont {{P. C. Bressloff}}},\
  }\href@noop {} {\bibfield  {journal} {\bibinfo  {journal} {J Phys A}\
  }\textbf {\bibinfo {volume} {50}},\ \bibinfo {pages} {133001} (\bibinfo
  {year} {2017})}\BibitemShut {NoStop}%
\bibitem [{\citenamefont {{M. B. Elowitz, A. J. Levine, E.D. Siggia and P. S.
  Swain}}(2002)}]{Elowitzetal}%
  \BibitemOpen
  \bibfield  {author} {\bibinfo {author} {\bibnamefont {{M. B. Elowitz, A. J.
  Levine, E.D. Siggia and P. S. Swain}}},\ }\href@noop {} {\bibfield  {journal}
  {\bibinfo  {journal} {Science}\ }\textbf {\bibinfo {volume} {297}},\ \bibinfo
  {pages} {1183} (\bibinfo {year} {2002})}\BibitemShut {NoStop}%
\bibitem [{\citenamefont {{A. Eldar and M. B. Elowitz}}(2010)}]{Eldar}%
  \BibitemOpen
  \bibfield  {author} {\bibinfo {author} {\bibnamefont {{A. Eldar and M. B.
  Elowitz}}},\ }\href@noop {} {\bibfield  {journal} {\bibinfo  {journal}
  {Nature}\ }\textbf {\bibinfo {volume} {467}},\ \bibinfo {pages} {167}
  (\bibinfo {year} {2010})}\BibitemShut {NoStop}%
\bibitem [{\citenamefont {{T. M. Norman, N. D. Lord, J. Paulsson and R.
  Losick}}(2015)}]{Normanetal}%
  \BibitemOpen
  \bibfield  {author} {\bibinfo {author} {\bibnamefont {{T. M. Norman, N. D.
  Lord, J. Paulsson and R. Losick}}},\ }\href@noop {} {\bibfield  {journal}
  {\bibinfo  {journal} {Ann Rev Microbiol}\ }\textbf {\bibinfo {volume} {69}},\
  \bibinfo {pages} {381} (\bibinfo {year} {2015})}\BibitemShut {NoStop}%
\bibitem [{\citenamefont {{C. W. Gardiner}}(2009)}]{gardiner}%
  \BibitemOpen
  \bibfield  {author} {\bibinfo {author} {\bibnamefont {{C. W. Gardiner}}},\
  }\href@noop {} {\emph {\bibinfo {title} {Stochastic Methods: a Handbook for
  the Natural and Social Sciences}}}\ (\bibinfo  {publisher} {Springer},\
  \bibinfo {year} {2009})\BibitemShut {NoStop}%
\bibitem [{\citenamefont {{A. M. Walczak, A. Mugler and C. H.
  Wiggins}}(2012)}]{walczak}%
  \BibitemOpen
  \bibfield  {author} {\bibinfo {author} {\bibnamefont {{A. M. Walczak, A.
  Mugler and C. H. Wiggins}}},\ }in\ \href@noop {} {\emph {\bibinfo {booktitle}
  {Computational Modeling of Signaling Networks}}},\ \bibinfo {editor} {edited
  by\ \bibinfo {editor} {\bibfnamefont {X.}~\bibnamefont {Liu}}\ and\ \bibinfo
  {editor} {\bibfnamefont {M.~D.}\ \bibnamefont {Betterton}}}\ (\bibinfo
  {publisher} {Springer},\ \bibinfo {year} {2012})\ pp.\ \bibinfo {pages}
  {273--322}\BibitemShut {NoStop}%
\bibitem [{\citenamefont {{V. Shahrezaei and P. S.
  Swain}}(2008)}]{ShahrezaeiSwain2008}%
  \BibitemOpen
  \bibfield  {author} {\bibinfo {author} {\bibnamefont {{V. Shahrezaei and P.
  S. Swain}}},\ }\href@noop {} {\bibfield  {journal} {\bibinfo  {journal} {Proc
  Nat Acad Sci (USA)}\ }\textbf {\bibinfo {volume} {105}},\ \bibinfo {pages}
  {17256} (\bibinfo {year} {2008})}\BibitemShut {NoStop}%
\bibitem [{\citenamefont {{R. S. Maier and D. L. Stein}}(1993)}]{MaierStein92}%
  \BibitemOpen
  \bibfield  {author} {\bibinfo {author} {\bibnamefont {{R. S. Maier and D. L.
  Stein}}},\ }\href@noop {} {\bibfield  {journal} {\bibinfo  {journal} {Phys
  Rev E}\ }\textbf {\bibinfo {volume} {48}},\ \bibinfo {pages} {931} (\bibinfo
  {year} {1993})}\BibitemShut {NoStop}%
\bibitem [{\citenamefont {{H. Eyring}}(1935)}]{eyring1935activated}%
  \BibitemOpen
  \bibfield  {author} {\bibinfo {author} {\bibnamefont {{H. Eyring}}},\
  }\href@noop {} {\bibfield  {journal} {\bibinfo  {journal} {J Chem Phys}\
  }\textbf {\bibinfo {volume} {3}},\ \bibinfo {pages} {107} (\bibinfo {year}
  {1935})}\BibitemShut {NoStop}%
\bibitem [{\citenamefont {{C. C. Chow and M. A. Buice}}(2015)}]{chow}%
  \BibitemOpen
  \bibfield  {author} {\bibinfo {author} {\bibnamefont {{C. C. Chow and M. A.
  Buice}}},\ }\href@noop {} {\bibfield  {journal} {\bibinfo  {journal} {J Math
  Neuro}\ }\textbf {\bibinfo {volume} {5}},\ \bibinfo {pages} {8} (\bibinfo
  {year} {2015})}\BibitemShut {NoStop}%
\bibitem [{\citenamefont {{A. M. Assaf, E. Roberts and Z.
  Luthey-Schulten}}(2011)}]{assafPRL2011}%
  \BibitemOpen
  \bibfield  {author} {\bibinfo {author} {\bibnamefont {{A. M. Assaf, E.
  Roberts and Z. Luthey-Schulten}}},\ }\href@noop {} {\bibfield  {journal}
  {\bibinfo  {journal} {Phys Rev Lett}\ }\textbf {\bibinfo {volume} {106}},\
  \bibinfo {pages} {248102} (\bibinfo {year} {2011})}\BibitemShut {NoStop}%
\bibitem [{\citenamefont {{N. Friedman, L. Cai and X. S. Xie}}(2006)}]{fried}%
  \BibitemOpen
  \bibfield  {author} {\bibinfo {author} {\bibnamefont {{N. Friedman, L. Cai
  and X. S. Xie}}},\ }\href@noop {} {\bibfield  {journal} {\bibinfo  {journal}
  {Phys Rev Lett}\ }\textbf {\bibinfo {volume} {97}},\ \bibinfo {pages}
  {168302} (\bibinfo {year} {2006})}\BibitemShut {NoStop}%
\bibitem [{\citenamefont {{L. Cai, N. Friedman and X. S. Xie}}(2006)}]{caifri}%
  \BibitemOpen
  \bibfield  {author} {\bibinfo {author} {\bibnamefont {{L. Cai, N. Friedman
  and X. S. Xie}}},\ }\href@noop {} {\bibfield  {journal} {\bibinfo  {journal}
  {Nature(London)}\ }\textbf {\bibinfo {volume} {440}},\ \bibinfo {pages} {358}
  (\bibinfo {year} {2006})}\BibitemShut {NoStop}%
\bibitem [{\citenamefont {{J. Yu, J. Xiao, Jie, X. Ren, K. Lao and X. S.
  Xie}}(2006)}]{Yu}%
  \BibitemOpen
  \bibfield  {author} {\bibinfo {author} {\bibnamefont {{J. Yu, J. Xiao, Jie,
  X. Ren, K. Lao and X. S. Xie}}},\ }\href@noop {} {\bibfield  {journal}
  {\bibinfo  {journal} {Science}\ }\textbf {\bibinfo {volume} {311}},\ \bibinfo
  {pages} {1600} (\bibinfo {year} {2006})}\BibitemShut {NoStop}%
\bibitem [{\citenamefont {{P. C. Bressloff}}(2010)}]{bress2}%
  \BibitemOpen
  \bibfield  {author} {\bibinfo {author} {\bibnamefont {{P. C. Bressloff}}},\
  }\href@noop {} {\bibfield  {journal} {\bibinfo  {journal} {Phys Rev E}\
  }\textbf {\bibinfo {volume} {82}},\ \bibinfo {pages} {051903} (\bibinfo
  {year} {2010})}\BibitemShut {NoStop}%
\bibitem [{\citenamefont {{D. T. Gillespie}}(1977)}]{Gillespie}%
  \BibitemOpen
  \bibfield  {author} {\bibinfo {author} {\bibnamefont {{D. T. Gillespie}}},\
  }\href@noop {} {\bibfield  {journal} {\bibinfo  {journal} {J Phys Chem}\
  }\textbf {\bibinfo {volume} {81}},\ \bibinfo {pages} {2340} (\bibinfo {year}
  {1977})}\BibitemShut {NoStop}%
\bibitem [{\citenamefont {{W. H. Press, S. A. Teukolsky, W. T. Vetterling and
  B. P. Flannery}}(2007)}]{press2007numerical}%
  \BibitemOpen
  \bibfield  {author} {\bibinfo {author} {\bibnamefont {{W. H. Press, S. A.
  Teukolsky, W. T. Vetterling and B. P. Flannery}}},\ }\href@noop {} {\emph
  {\bibinfo {title} {Numerical Recipes: The Art of Scientific Computing (3rd
  edition)}}}\ (\bibinfo  {publisher} {Cambridge University Press},\ \bibinfo
  {year} {2007})\BibitemShut {NoStop}%
\bibitem [{\citenamefont {{D. M. Roma, R. A. O'Flanagan, A. E. Ruckenstein, A.
  M. Sengupta and R. Mukhopadhyay}}(2005)}]{roma2005optimal}%
  \BibitemOpen
  \bibfield  {author} {\bibinfo {author} {\bibnamefont {{D. M. Roma, R. A.
  O'Flanagan, A. E. Ruckenstein, A. M. Sengupta and R. Mukhopadhyay}}},\
  }\href@noop {} {\bibfield  {journal} {\bibinfo  {journal} {Phys Rev E}\
  }\textbf {\bibinfo {volume} {71}},\ \bibinfo {pages} {011902} (\bibinfo
  {year} {2005})}\BibitemShut {NoStop}%
\bibitem [{\citenamefont {{R. S. Maier and D. L. Stein}}(1997)}]{maierstein1}%
  \BibitemOpen
  \bibfield  {author} {\bibinfo {author} {\bibnamefont {{R. S. Maier and D. L.
  Stein}}},\ }\href@noop {} {\bibfield  {journal} {\bibinfo  {journal} {SIAM J
  Appl Math}\ }\textbf {\bibinfo {volume} {57}},\ \bibinfo {pages} {752}
  (\bibinfo {year} {1997})}\BibitemShut {NoStop}%
\bibitem [{\citenamefont {{N. Yildirim and M. C.
  Mackey}}(2003)}]{YildirimMackey}%
  \BibitemOpen
  \bibfield  {author} {\bibinfo {author} {\bibnamefont {{N. Yildirim and M. C.
  Mackey}}},\ }\href@noop {} {\bibfield  {journal} {\bibinfo  {journal}
  {Biophys J}\ }\textbf {\bibinfo {volume} {84}},\ \bibinfo {pages} {2841}
  (\bibinfo {year} {2003})}\BibitemShut {NoStop}%
\bibitem [{\citenamefont {{N. Yildirim, M. Santillan, D. Horike and M. C.
  Mackey}}(2004)}]{Yildirimetal}%
  \BibitemOpen
  \bibfield  {author} {\bibinfo {author} {\bibnamefont {{N. Yildirim, M.
  Santillan, D. Horike and M. C. Mackey}}},\ }\href@noop {} {\bibfield
  {journal} {\bibinfo  {journal} {Chaos}\ }\textbf {\bibinfo {volume} {14}},\
  \bibinfo {pages} {279} (\bibinfo {year} {2004})}\BibitemShut {NoStop}%
\bibitem [{\citenamefont {{P. M. Bhogale, R. A. Sorg, J.-W. Veening and J.
  Berg}}(2014)}]{Bhogale}%
  \BibitemOpen
  \bibfield  {author} {\bibinfo {author} {\bibnamefont {{P. M. Bhogale, R. A.
  Sorg, J.-W. Veening and J. Berg}}},\ }\href@noop {} {\bibfield  {journal}
  {\bibinfo  {journal} {Nucleic Acids Res}\ }\textbf {\bibinfo {volume} {42}},\
  \bibinfo {pages} {11321} (\bibinfo {year} {2014})}\BibitemShut {NoStop}%
\bibitem [{\citenamefont {{V. Chickarmane, V. Olariu and C.
  Peterson}}(2012)}]{Chickarmane}%
  \BibitemOpen
  \bibfield  {author} {\bibinfo {author} {\bibnamefont {{V. Chickarmane, V.
  Olariu and C. Peterson}}},\ }\href@noop {} {\bibfield  {journal} {\bibinfo
  {journal} {BMC Systems Biol}\ }\textbf {\bibinfo {volume} {6}},\ \bibinfo
  {pages} {98} (\bibinfo {year} {2012})}\BibitemShut {NoStop}%
\bibitem [{Note1()}]{Note1}%
  \BibitemOpen
  \bibinfo {note} {Https://home.icts.res.in/~abhi/notes/kram.pdf}\BibitemShut
  {NoStop}%
\bibitem [{\citenamefont {{C. Escudero and A. Kamenev}}(2009)}]{escuder}%
  \BibitemOpen
  \bibfield  {author} {\bibinfo {author} {\bibnamefont {{C. Escudero and A.
  Kamenev}}},\ }\href@noop {} {\bibfield  {journal} {\bibinfo  {journal} {Phys
  Rev E}\ }\textbf {\bibinfo {volume} {79}},\ \bibinfo {pages} {041149}
  (\bibinfo {year} {2009})}\BibitemShut {NoStop}%
\end{thebibliography}%

\end{document}